\documentclass[12pt,a4paper,pdflatex]{article}%for arXiv (dvipdfmx poses problems)
\usepackage{amsmath,amsfonts,amsthm,amssymb,graphicx}
\usepackage{cite}
\usepackage{tikz}
\usepackage{ytableau}
\allowdisplaybreaks[4]
\numberwithin{equation}{section} 
\newtheorem{theorem}{Theorem}[section]
\newtheorem{lemma}[theorem]{Lemma}
\newtheorem{proposition}[theorem]{Proposition}

\usepackage[body={16cm,23cm}]{geometry}
\newcommand{\Bs}{{\mathsf B}}
\newcommand{\Cs}{{\mathsf C}}
\newcommand{\Ds}{{\mathsf D}}
\begin{document}
\title{
Character Formulas for Kirillov-Reshetikhin Modules via Folding of Supercharacters of $\mathfrak{gl}(M|N)$. 
}
\author{Zengo Tsuboi
%\footnote{
%E-mail: ztsuboi$\bullet$yahoo.co.jp
%}
\\[8pt]
{\sl Osaka Central Advanced Mathematical Institute (OCAMI),}
\\
{\sl Osaka Metropolitan University,}
\\
{\sl 3-3-138 Sugimoto, Sumiyoshi-ku Osaka 558-8585, Japan}
} 
%\date{}
\maketitle
%%%%%%%%%%%%%%%%%%%%%%%%%%%%%%%%
\begin{abstract}
We derive decomposition formulas for supercharacters of quantum affine orthosymplectic superalgebras and twisted quantum 
affine superalgebras into supercharacters of their finite-type quantum sub-superalgebras, by employing Cauchy-type identities for supersymmetric Schur functions.
These formulas are obtained via a folding (reduction) procedure applied to the supercharacters of the finite-dimensional general linear Lie superalgebra $\mathfrak{gl}(M|N)$. As a special case, 
our results provide explicit character formulas for a class of
Kirillov-Reshetikhin modules of quantum affine algebras (and their
Yangian counterparts), thereby proving a previously proposed conjecture derived from Bethe ansatz analysis.
\end{abstract}
%%%%%%%
\\[5pt]
Journal reference: Nuclear Physics B 1025 (2026) 117400
\\[5pt]
DOI: https://doi.org/10.1016/j.nuclphysb.2026.117400
%%%%%%%%%%%%%%%%%%%%%%%%%%%%%%%%%%%%%%%%%%%%%%%%%%%%%%%%%%%%%%%%%%%%
\section{Introduction}
Quantum affine algebras have been intensively studied over the past decades, both for their intrinsic algebraic structure and for their applications in mathematical physics. Their representation theory exhibits a remarkable richness that has inspired extensive research in integrable systems, combinatorics, geometry, and related fields. 
Within this framework, the Kirillov-Reshetikhin (KR) modules play a central role, as their characters appear in functional relations such as the Q-system \cite{KR90,Ki89,HKOTY98,HKOTT01} and serve as key elements in elucidating the representation-theoretic and symmetry structures of quantum integrable systems.

In the context of the Bethe ansatz, 
we conjectured in \cite{T23} (see also \cite{T11,T21})---based on computational experiments---that the characters of KR modules for a large class of 
quantum affine algebras can be described through a folding (or reduction) procedure applied to supercharacters of the 
 finite-dimensional general linear Lie superalgebra $\mathfrak{gl}(M|N)$. In this paper, we establish this conjecture by making use of Cauchy-type relations for supercharacters of $\mathfrak{gl}(M|N)$ and $\mathfrak{osp}(M|N)$ 
 (see, Theorem \ref{main-th}). These relations allow us to connect different families of supercharacters and to control their combinatorial structure in a way suitable for proving the folding identity.

A general difficulty in the theory of KR modules is that for quantum affine algebras $U_q(\mathfrak{g}_{\mathrm{aff}})$ of non-type A, evaluation homomorphisms from $U_q(\mathfrak{g}_{\mathrm{aff}})$ to the quantum finite-type subalgebra $U_q(\mathfrak{g})$ are generally absent. 
As a consequence, KR modules cannot be realized directly as 
evaluation modules. When restricted to 
the finite-type quantum group $U_q(\mathfrak{g})$, such a module typically decomposes as
\begin{align}
W^{\mathrm{KR}}_{\Lambda} \cong V({\Lambda}) \oplus \bigoplus_{\Lambda^{\prime} < \Lambda} m_{\Lambda^{\prime}}  V(\Lambda^{\prime}),
\label{KRdecom}
\end{align}
where $V(\Lambda)$ denotes the distinguished highest weight 
module with the highest weight $\Lambda$, 
and the remaining terms correspond to lower 
highest weight modules 
$V(\Lambda^{\prime})$ appearing with multiplicities $m_{\Lambda^{\prime}}$.
The highest weight module $V(\Lambda)$ is \emph{rectangular} 
in the sense that, at least in the non-superalgebraic case, 
the highest weight takes the form 
\begin{align}
 \Lambda= m \Lambda_{a}, \label{hw-nonsuper}
\end{align} where $\Lambda_{a}$ is the $a$-th fundamental 
weight ($a \in \{1,2,\dots , \mathrm{rank}(\mathfrak{g})\}$) and  
$m \in {\mathbb Z}_{\ge 1}$. 
In many instances, such a highest weight module can be conveniently represented by a rectangular Young diagram.
We denote $W^{(a)}_{m}=W^{\mathrm{KR}}_{m \Lambda_{a}}$. 
Strictly speaking, KR modules as irreducible 
$U_q(\mathfrak{g}_{\mathrm{aff}})$-modules also depend on a spectral parameter and are characterized by their Drinfeld polynomials. In the present paper, however, we suppress this dependence, since we consider only their (super)characters, which are independent of the spectral parameter.

The decomposition \eqref{KRdecom} reflects the fact that KR modules, while carrying a natural highest weight, do not, in general, remain irreducible when restricted to the finite-type subalgebra.
For $(\mathfrak{g}_{\mathrm{aff}},\mathfrak{g})=
(\mathfrak{gl}(M|N)^{(1)},\mathfrak{gl}(M|N))$, however, the situation is different 
in that $m_{\lambda^{\prime}}=0$ for any $\lambda^{\prime}<\lambda$ 
in \eqref{KRdecom}: the algebra 
$U_q(\mathfrak{gl}(M|N)^{(1)})$ does admit evaluation maps, so that KR modules can be constructed explicitly as evaluation modules of finite-dimensional $U_q(\mathfrak{gl}(M|N))$-modules, and their supercharacters coincide with of those of $\mathfrak{gl}(M|N)$. The main achievement of this paper is to show that such a description can be extended beyond type A. Even when evaluation homomorphisms are absent, a large class of 
KR characters can still be obtained as specializations of supercharacters of 
$\mathfrak{gl}(M|N)$ via an explicit folding construction.

This paper is organized as follows. In Section 2, we briefly summarize the relation between 
 Young diagrams and highest weights of representations of
 finite dimensional Lie superalgebras. These are necessary for labeling the 
 (super)character formulas in the subsequent sections.  
In Section 3, we review Cauchy-type identities on
 generating functions of supersymmetric Schur functions in detail
 and reformulate them to address our problems.
In Section 4, we specialize these identities and obtain objects that can be naturally interpreted as supercharacters of representations of the quantum affine superalgebras
$U_q(\mathfrak{osp}(2r+1|2s)^{(1)})$, $U_q(\mathfrak{sl}(2r|2s+1)^{(2)})$, 
$U_q(\mathfrak{sl}(2r+1|2s)^{(2)})$, $U_q(\mathfrak{sl}(2r|2s)^{(2)})$, 
$U_q(\mathfrak{osp}(2r|2s)^{(1)})$, and
$U_q(\mathfrak{osp}(2r|2s)^{(2)})$ (and, for the untwisted case, also their Yangian counterparts). This interpretation is motivated by the observation that these objects arise as limiting expressions of transfer-matrix eigenvalues of quantum integrable superspin chains, as discussed in our previous work \cite{T23}. 
These supercharacters are not necessarily irreducible in general, especially for the $U_q(\mathfrak{osp}(2r|2s)^{(1)})$ case, but 
upon the specialization $s=0$ (or $r=0$), we 
reproduce a large class of irreducible 
characters of the Kirillov-Reshetikhin modules  
of $U_q(\mathfrak{so}(2r+1)^{(1)})$,  $U_q(\mathfrak{sl}(2r+1)^{(2)})$
,  $U_q(\mathfrak{sl}(2r)^{(2)})$, $U_q(\mathfrak{so}(2r)^{(1)})$,
 $U_q(\mathfrak{sp}(2r)^{(1)})$, and $U_q(\mathfrak{so}(2r)^{(2)})$.
  It is known that there are correspondences between representations of untwisted quantum affine (super)algebras 
 and twisted quantum affine (super)algebras 
(cf. \cite{Z97,XZ16}). 
 Our work exemplifies the existence of this type of nontrivial correspondence, mediated through identities among supercharacters.
 It should be noted that in this paper we focus on tensor-like (spin-even) representations, for which the relevant KR modules correspond to rectangular highest weights. The case of spinor-like (spin-odd) representations requires a separate treatment: as discussed in our earlier works \cite{T21,T23}, one needs to consider asymptotic limits of supercharacters of typical $\mathfrak{gl}(M|N)$-modules, together with an appropriate folding procedure. 
 Section 5 is devoted to concluding remarks.

To summarize, the present work confirms our earlier conjecture  \cite{T23} and shows that KR characters can be realized as folded supercharacters of $\mathfrak{gl}(M|N)$. Beyond providing explicit character formulas, this approach offers a unified viewpoint across different types of quantum affine algebras and their super counterparts, thereby clarifying the structural role of folding and specialization in the representation theory of quantum affine algebras.

As this paper proves the conjectures introduced in \cite{T23}, some portions of the text overlap with Section 2 and Appendix B in \cite{T23}.

%%%%%%%%%%%%%%%%%%%%%
%%%%%%%%%%%%%%%%%%%%%%%%%%%%%%%

{\bf Notation:}
We assume that the deformation parameter $q$ of the quantum groups 
is not a root of unity. 
Let ${\mathcal P}$ be the set of partitions: 
${\mathcal P}=\{\lambda =(\lambda_{1},\lambda_{2},\dots ) \in ({\mathbb Z_{\ge 0}})^{\infty}| \lambda_{1} \ge \lambda_{2} \ge \dots \}$, 
$\lambda^{\prime} =(\lambda_{1}^{\prime},\lambda_{2}^{\prime},\dots ) $ be the conjugate of $\lambda =(\lambda_{1},\lambda_{2},\dots ) \in {\mathcal P}$, where $\lambda_{i}^{\prime}=\text{Card}\{j | \lambda_{j} \ge i\}$.
The length $l=l(\lambda)$ of $\lambda =(\lambda_{1},\lambda_{2},\dots ) \in {\mathcal P}$ is determined by 
$\lambda_{l}>0$, $\lambda_{l+1}=0$. Thus $l(\lambda)=\lambda_{1}^{\prime}$. The size of $\lambda $ is defined by $|\lambda|=\sum_{j=1}^{\infty}\lambda_{j}$. 
We will also use subsets of ${\mathcal P}$: 
${\mathcal P}^{+}=\{\lambda =(\lambda_{1},\lambda_{2},\dots ) \in {\mathcal P}|\lambda_{i} 
\in 2{\mathbb Z} \text{ for any } i\}$ and 
${\mathcal P}^{-}=\{\lambda =(\lambda_{1},\lambda_{2},\dots ) \in {\mathcal P}|\lambda^{\prime }_{i} 
\in 2{\mathbb Z} \text{ for any } i\}$.
%%%%%%%%%%%%%%%%%%

\section{Lie superalgebras}
\label{Liesuper}
In this section, we summarize the simple roots and highest weights of finite dimensional Lie superalgebras that are needed for labeling the quantities discussed in what follows. 
For Young diagrams and labeling of representations of orthosymplectic Lie superalgebras, see \cite{FJ84, MSS85}. 
For general references on Lie superalgebras\footnote{For quantum affine superalgebras, see \cite{Yamane99, KT94}.}, see, for example, \cite{Kac78, FSS89, FSS20, CW12}.

Each simple root $\alpha$ of a basic Lie superalgebra carries a grading (parity) 
$p_{\alpha} \in \{1,-1\}$.  
The root $\alpha$ is called an {\em even} (bosonic) root if $p_{\alpha}=1$, and an 
{\em odd} (fermionic) root if $p_{\alpha}=-1$.  
There is an inner product $(\cdot \mid \cdot)$ on the root system.  
Unlike ordinary Lie algebras, Lie superalgebras admit several inequivalent simple root systems.  
Among them, the simplest is the {\em distinguished} simple root system, which contains exactly one odd root.  
In this paper, we work exclusively with the distinguished simple root system for each superalgebra.

One can draw a Dynkin diagram for any simple root system.  
To each simple root $\alpha$, one assigns one of the following three types of nodes:

\begin{tikzpicture}[x=1.1pt,y=1.1pt,line width=0.8pt]
\draw (0,0) node[left] {white node};
\draw (7,0) circle (6);
\draw (15,-1) node[right] {if $(\alpha \mid \alpha)\neq 0$ and $p_{\alpha}=1$};
\end{tikzpicture}

\begin{tikzpicture}[x=1.1pt,y=1.1pt,line width=0.8pt]
\draw (0,0) node[left] {black node};
\fill (7,0) circle (6);
\draw (15,0) node[right] {if $(\alpha \mid \alpha)\neq 0$ and $p_{\alpha}=-1$};
\end{tikzpicture}

\begin{tikzpicture}[x=1.1pt,y=1.1pt,line width=0.8pt]
\draw (0,0) node[left] {gray node};
\draw (7,0) circle (6);
\draw (7-4.24,-4.24)--(7+4.24,4.24);
\draw (7-4.24,4.24)--(7+4.24,-4.24);
\draw (15,0) node[right] {if $(\alpha \mid \alpha)=0$ and $p_{\alpha}=-1$};
\end{tikzpicture}

The black node appears, for example, in the Dynkin diagrams of 
$\mathfrak{osp}(2r+1 \mid 2s)$.

Let $\{\varepsilon_i\} \sqcup \{\delta_i\}$ be a basis of the dual space of a Cartan subalgebra of a Lie superalgebra, equipped with the bilinear form
\[
 (\varepsilon_i \mid \varepsilon_j)=\delta_{ij},\qquad
 (\varepsilon_i \mid \delta_j)=(\delta_j \mid \varepsilon_i)=0,\qquad
 (\delta_i \mid \delta_j)=-\delta_{ij}.
\]
Using this notation, we describe in the following the simple root systems of the general linear and the orthosymplectic Lie superalgebras.

%%%%%%%%%%%%%%%%%%%%%%%%%%%%%%%
%\subsubsection{Simple root systems and highest weights}
\paragraph{Type A}
The distinguished simple root system of $\mathfrak{gl}(M|N)$ is given by 
\begin{align}
\begin{split}
\alpha_{i}&=\varepsilon_{i}-\varepsilon_{i+1} \quad \text{for} \quad i \in \{1,2,\dots , M-1\},
\\
\alpha_{M}&=\varepsilon_{M}-\delta_{1},
\\
\alpha_{i+M}&=\delta_{i}-\delta_{i+1} \quad \text{for} \quad i \in \{1,2,\dots , N-1\},
\end{split}
\label{rootAd}
\end{align}
and the corresponding Dynkin diagram is  

%%%
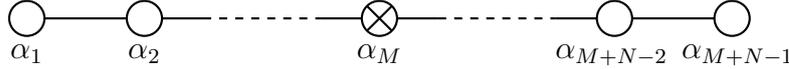
\begin{tikzpicture}[x=1.1pt,y=1.1pt,line width=0.8pt]
\draw  (0,0) circle (6);
\draw (6,0) -- (34,0);
\draw  (40,0) circle (6);
\draw (46,0) -- (60,0);
\draw[dashed] (60,0) -- (100,0);
\draw (100,0) -- (114,0);
\draw  (120,0) circle (6);
\draw (120-4.24,-4.24) -- (120+4.24,4.24);
\draw (120-4.24,4.24) -- (120+4.24,-4.24);
\draw (126,0) -- (140,0);
\draw[dashed] (140,0) -- (180,0);
\draw (180,0) -- (194,0);
\draw (200,0) circle (6);
\draw (206,0) -- (234,0);
\draw (240,0) circle (6);
\draw (0,-5) node[below] {$\alpha_{1}$};
\draw (40,-5) node[below] {$\alpha_{2}$};
\draw (120,-5) node[below] {$\alpha_{M}$};
\draw (199,-5) node[below] {$\alpha_{M+N-2}$};
\draw (242,-5) node[below] {$\alpha_{M+N-1}$};
\end{tikzpicture} 
%

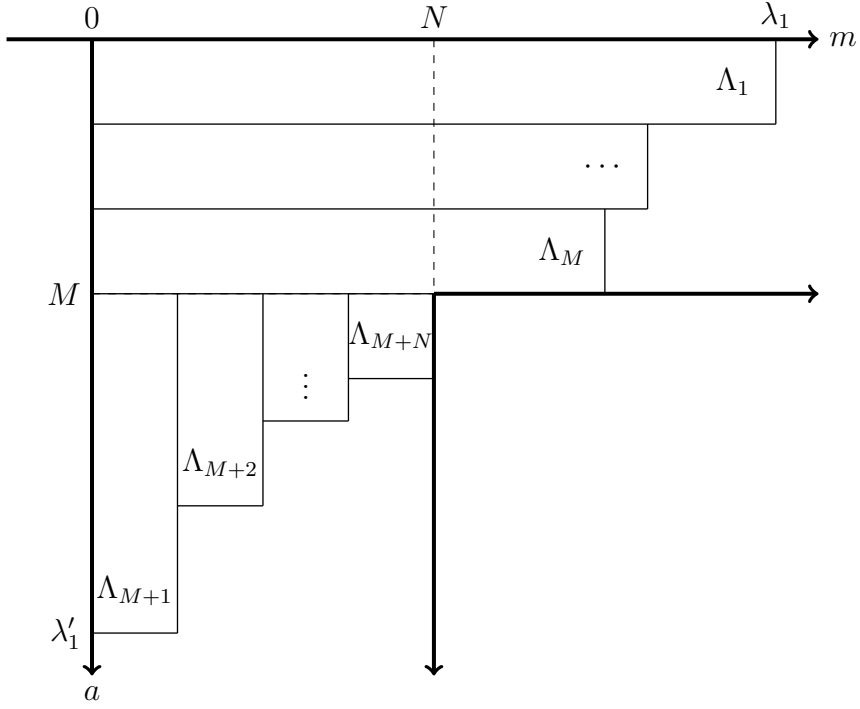
\begin{figure}
\centering
\begin{tikzpicture}[x=1.6pt,y=1.6pt]
%,line width=1.8pt]
\draw[-to,line width=1.5pt]  (-20,0) -- (170,0);
\draw[-to,line width=1.5pt] (80,-60) -- (170,-60);
\draw[-to,line width=1.5pt] (0,0) -- (0,-150);
\draw[-to,line width=1.5pt] (80,-60) -- (80,-150);
\draw[dashed]  (0,-60) -- (80,-60);
\draw[dashed]  (80,0) -- (80,-60);
\draw[line width=0.5pt] (0,-20) -- (160,-20);
\draw[line width=0.5pt] (160,0) -- (160,-20);
\draw[line width=0.5pt] (0,-40) -- (130,-40);
\draw[line width=0.5pt] (130,-20) -- (130,-40);
\draw[line width=0.5pt] (0,-60) -- (120,-60);
\draw[line width=0.5pt] (120,-40) -- (120,-60);
\draw[line width=0.5pt] (60,-80) -- (80,-80);
\draw[line width=0.5pt] (60,-60) -- (60,-90);
\draw[line width=0.5pt] (40,-90) -- (60,-90);
\draw[line width=0.5pt] (40,-60) -- (40,-110);
\draw[line width=0.5pt] (20,-110) -- (40,-110);
\draw[line width=0.5pt] (20,-60) -- (20,-140);
\draw[line width=0.5pt] (0,-140) -- (20,-140);
%\draw[-to,line width=1pt] (32) -- node[midway,left]{$F_{1}^{[2]}$} (22); 
%\draw[dashed] (60,0) -- (140,0);
%\draw[double,double distance=3pt] (200,0) -- (240-6,0);
%
%\draw (240,0) circle (6);
%
\draw (0,0) node[above] {$0$};
\draw (80,0) node[above] {$N$};
\draw (160,0) node[above] {$\lambda_{1}$};
\draw (0,-60) node[left] {$M$};
\draw (0,-140) node[left] {$\lambda_{1}^{\prime}$};
\draw (150,-10) node {$\Lambda_{1}$};
\draw (120,-30) node {$ \cdots $};
\draw (110,-50) node {$ \Lambda_{M}$};
\draw (10,-130) node {$ \Lambda_{M+1}$};
\draw (30,-100) node {$ \Lambda_{M+2}$};
\draw (50,-80) node {$ \vdots $};
\draw (70,-70) node {$ \Lambda_{M+N}$};
\draw (0,-150) node[below] {$a$};
\draw (170,0) node[right] {$m$};
\end{tikzpicture} 
\caption{$[M,N]$-hook: the Young diagram $\lambda $ is related to the highest weight \eqref{HW-A} by \eqref{YW-A}.}
\label{MN-hookA}
\end{figure}

%%%%%%
Let $V(\Lambda )$ be the irreducible module of $\mathfrak{gl}(M|N)$ with the highest weight
\begin{align}
\Lambda =\sum_{j=1}^{M} \Lambda_{j} \varepsilon_{j} +\sum_{j=1}^{N} \Lambda_{M+j} \delta_{j} ,
\label{HW-A}
\end{align}
where $\Lambda_{j} \in \mathbb{C}$.
The Kac-Dynkin labels $[b_{1},b_{2},\dots , b_{M+N-1}]$ of  $V(\Lambda )$ are given by 
\begin{align}
b_{j}=\Lambda_{j}-\Lambda_{j+1} \quad \text{for} \quad j \ne M, \qquad
b_{M}=\Lambda_{M}+\Lambda_{M+1}. \label{KD-A}
\end{align}
$V(\Lambda )$ is finite dimensional if $b_{j} \in \mathbb{Z}_{\ge 0}$ for $j \ne M$. 
In case $\Lambda_{j} \in \mathbb{Z}_{\ge 0} $, these parameters are related to 
an $[M,N]$-hook partition $\lambda =(\lambda_{1},\lambda_{2}, \dots )$, $\lambda_{1} \ge \lambda_{2} \ge \dots \ge 0$, 
$\lambda_{M+1} \le N$: 
\begin{align}
\Lambda_{j}=\lambda_{j} \quad \text{for} \quad j \in \{1,2,\dots , M \}, \quad 
\Lambda_{M+j}=\max \{ \lambda_{j}^{\prime} -M ,0\} 
\quad \text{for} \quad j \in \{1,2,\dots , N \} ,
\label{YW-A}
\end{align}
where $\lambda_{k}^{\prime}=\mathrm{Card}\{ j| \lambda_{j} \ge k \}$. The $[M,N]$-hook partition describes a Young 
diagram in the $[M,N]$-hook (see Figure \ref{MN-hookA}). 
We shall, in what follows, use the same symbol $\lambda$ 
to denote both a partition and the corresponding Young diagram.
%%%%%%%%%%%%
%%%%%%%%%%%%%%%%%%%%%%%%%%%%%%%%%%%%%%%%%%%%%

%%%%%%%%
\paragraph{Type B}
($r,s \in \mathbb{Z}_{\ge 0}$, $r+s \ge 1$) 

\subparagraph{The case $r>0$:}
The distinguished simple root system of $B(r|s)=\mathfrak{osp}(2r+1|2s)$ 
is given by
\begin{align}
\begin{split}
\alpha_{i}&=\delta_{i}-\delta_{i+1} \quad \text{for} \quad i \in \{1,2,\dots , s-1\},
\\
\alpha_{s}&=\delta_{s}-\varepsilon_{1},
\\
\alpha_{i+s}&=\varepsilon_{i}-\varepsilon_{i+1} \quad \text{for} \quad i \in \{1,2,\dots , r-1\},
\\
\alpha_{r+s}&=\varepsilon_{r},
\end{split}
\label{rootBd}
\end{align}

and the corresponding Dynkin diagram is 

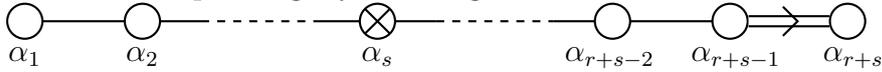
\begin{tikzpicture}[x=1.1pt,y=1.1pt,line width=0.8pt]
\draw  (0,0) circle (6);
\draw (6,0) -- (34,0);
\draw  (40,0) circle (6);
\draw (46,0) -- (60,0);
\draw[dashed] (60,0) -- (100,0);
\draw (100,0) -- (114,0);
\draw  (120,0) circle (6);
\draw (120-4.24,-4.24) -- (120+4.24,4.24);
\draw (120-4.24,4.24) -- (120+4.24,-4.24);
\draw (126,0) -- (140,0);
\draw[dashed] (140,0) -- (180,0);
\draw (180,0) -- (194,0);
\draw (200,0) circle (6);
\draw (206,0) -- (234,0);
\draw (240,0) circle (6);
\draw[double,double distance=3pt] (246.3,0) -- (273.8,0);
\draw (263-5,5) -- (263,0) -- (263-5,-5);
\draw (280,0) circle (6);
\draw (0,-5) node[below] {$\alpha_{1}$};
\draw (40,-5) node[below] {$\alpha_{2}$};
\draw (120,-5) node[below] {$\alpha_{s}$};
\draw (199,-5) node[below] {$\alpha_{r+s-2}$};
\draw (242,-5) node[below] {$\alpha_{r+s-1}$};
\draw (282,-5) node[below] {$\alpha_{r+s}$};
\end{tikzpicture} 

%%%%%%
Let $V(\Lambda )$ be the irreducible module of $\mathfrak{osp}(2r+1|2s)$ with the highest weight
\begin{align}
\Lambda =\sum_{j=1}^{s} \Lambda_{j} \delta_{j} +\sum_{j=1}^{r} \Lambda_{s+j} \varepsilon_{j} ,
\label{HW-B}
\end{align}
where $\Lambda_{j} \in \mathbb{C}$.
The Kac-Dynkin labels $[b_{1},b_{2},\dots , b_{r+s}]$ of  $V(\Lambda )$ are given by
\begin{align}
b_{j}=\Lambda_{j}-\Lambda_{j+1} \quad \text{for} \quad j \ne s,r+s, \qquad
b_{s}=\Lambda_{s}+\Lambda_{s+1}, \qquad b_{r+s}=2\Lambda_{r+s}. \label{KD-B}
\end{align}
$V(\Lambda )$ is finite dimensional if $b_{j} \in \mathbb{Z}_{\ge 0}$ for $j \ne s$, 
$c=b_{s}-b_{s+1}-b_{s+2}-\cdots -b_{r+s-1}-b_{r+s}/2 \in \mathbb{Z}_{\ge 0}$, and 
$b_{s+c+1}=b_{s+c+2}=\dots =b_{r+s}=0$ if $c < r$. 
In case $\Lambda_{j} \in \mathbb{Z}_{\ge 0} $, these parameters are related to 
an $[r,s]$-hook partition $\lambda =(\lambda_{1},\lambda_{2}, \dots )$, $\lambda_{1} \ge \lambda_{2} \ge \dots \ge 0$, 
$\lambda_{r+1} \le s$: 
\begin{align}
\Lambda_{j}=\lambda_{j}^{\prime} \quad \text{for} \quad j \in \{1,2,\dots , s \}, \quad 
\Lambda_{s+j}=\max \{ \lambda_{j} -s ,0\} 
\quad \text{for} \quad j \in \{1,2,\dots , r \} ,
\label{YW-B}
\end{align}
where $\lambda_{k}^{\prime}=\mathrm{Card}\{ j| \lambda_{j} \ge k \}$. 
We denote by $V_{\lambda}$ the module $V(\Lambda)$ 
defined in \eqref{YW-B}. 
The $[r,s]$-hook partition describes a Young 
diagram in the $[r,s]$-hook. This is embedded into the $[2r,2s+1]$-hook of $\mathfrak{gl}(2r|2s+1)$ (see Figure \ref{MN-hookB}). 
%%%%%%%%%%%%
\begin{figure}
\centering
\begin{tikzpicture}[x=1.6pt,y=1.6pt]
%,line width=1.8pt]
\draw[-to,line width=1.5pt]  (-20,0) -- (230,0);
\draw[-to,dashed, line width=1.5pt] (80,-60) -- (230,-60);
\draw[-to,line width=1.5pt] (180,-120) -- (230,-120);
\draw[-to,line width=1.5pt] (0,0) -- (0,-170);
\draw[-to,dashed,line width=1.5pt] (80,-60) -- (80,-170);
\draw[-to,line width=1.5pt] (180,-120) -- (180,-170);
\draw[dashed]  (0,-60) -- (80,-60);
\draw[dashed]  (0,-120) -- (180,-120);
\draw[dashed]  (80,0) -- (80,-60);
\draw[dashed]  (180,0) -- (180,-120);
\draw[line width=0.5pt] (80,-20) -- (210,-20);
\draw[line width=0.5pt] (210,0) -- (210,-20);
\draw[line width=0.5pt] (80,-40) -- (130,-40);
\draw[line width=0.5pt] (130,-20) -- (130,-40);
\draw[line width=0.5pt] (80,-60) -- (120,-60);
\draw[line width=0.5pt] (120,-40) -- (120,-60);
\draw[line width=0.5pt] (80,0) -- (80,-80);
\draw[line width=0.5pt] (60,-80) -- (80,-80);
\draw[line width=0.5pt] (60,0) -- (60,-90);
\draw[line width=0.5pt] (40,-90) -- (60,-90);
\draw[line width=0.5pt] (40,0) -- (40,-110);
\draw[line width=0.5pt] (20,-110) -- (40,-110);
\draw[line width=0.5pt] (20,0) -- (20,-140);
\draw[line width=0.5pt] (0,-140) -- (20,-140);
%\draw[-to,line width=1pt] (32) -- node[midway,left]{$F_{1}^{[2]}$} (22); 
%\draw[dashed] (60,0) -- (140,0);
%\draw[double,double distance=3pt] (200,0) -- (240-6,0);
%
%\draw (240,0) circle (6);
%
\draw (0,0) node[above] {$0$};
\draw (80,0) node[above] {$s$};
\draw (180,0) node[above] {$2s+1$};
\draw (210,0) node[above] {$\lambda_{1}$};
\draw (0,-60) node[left] {$r$};
\draw (0,-120) node[left] {$2r$};
\draw (0,-140) node[left] {$\lambda_{1}^{\prime}$};
\draw (200,-10) node {$\Lambda_{s+1}$};
\draw (120,-30) node {$ \cdots $};
\draw (110,-50) node {$ \Lambda_{s+r}$};
\draw (10,-130) node {$ \Lambda_{1}$};
\draw (30,-100) node {$ \Lambda_{2}$};
\draw (50,-80) node {$ \vdots $};
\draw (70,-70) node {$ \Lambda_{s}$};
\draw (0,-170) node[below] {$a$};
\draw (230,0) node[right] {$m$};
\end{tikzpicture} 
\caption{$[r,s]$-hook in $[2r,2s+1]$-hook: the Young diagram $\lambda $ is related to the highest weight \eqref{HW-B} by \eqref{YW-B}.}
\label{MN-hookB}
\end{figure}
%%%%%%%%%%%%%%%%%%%%%%%%%%%%%%%%%%%%%%%%%%%%%

%%%%%%%%%%
%%%%%%%%%%%%%%

\subparagraph{The case $r=0$:} 
The distinguished simple root system of $B(0|s)=\mathfrak{osp}(1|2s)$ is given by 
\begin{align}
\begin{split}
\alpha_{i}&=\delta_{i}-\delta_{i+1} \quad \text{for} \quad i \in \{1,2,\dots , s-1\},
\\
\alpha_{s}&=\delta_{s},
\end{split}
\label{rootBd0}
\end{align}

and the corresponding Dynkin diagram has the form

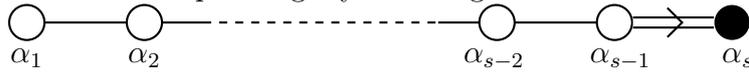
\begin{tikzpicture}[x=1.1pt,y=1.1pt,line width=0.8pt]
\draw  (0,0) circle (6);
\draw (6,0) -- (34,0);
\draw  (40,0) circle (6);
\draw (46,0) -- (60,0);
\draw[dashed] (60,0) -- (140,0);
\draw (140,0) -- (154,0);
\draw (160,0) circle (6);
\draw (166,0) -- (194,0);
\draw (200,0) circle (6);
\draw[double,double distance=3pt] (206.3,0) -- (233.8,0);
\draw (223-5,5) -- (223,0) -- (223-5,-5);
\fill (240,0) circle (6);
\draw (0,-5) node[below] {$\alpha_{1}$};
\draw (40,-5) node[below] {$\alpha_{2}$};
\draw (159,-5) node[below] {$\alpha_{s-2}$};
\draw (202,-5) node[below] {$\alpha_{s-1}$};
\draw (242,-5) node[below] {$\alpha_{s}$};
\end{tikzpicture} 

%%%%%%
Let $V(\Lambda )$ be the irreducible module of $\mathfrak{osp}(1|2s)$ with the highest weight
\begin{align}
\Lambda =\sum_{j=1}^{s} \Lambda_{j} \delta_{j}  ,
\label{HW-B0}
\end{align}
where $\Lambda_{j} \in \mathbb{C}$.
The Kac-Dynkin labels $[b_{1},b_{2},\dots , b_{s}]$ of  $V(\Lambda )$ are given by 
\begin{align}
b_{j}=\Lambda_{j}-\Lambda_{j+1} \quad \text{for} \quad j \ne s,  \qquad b_{s}=2\Lambda_{s}. \label{KD-B0}
\end{align}
$V(\Lambda )$ is finite dimensional if $b_{j} \in \mathbb{Z}_{\ge 0}$ for $j \ne s$, 
$c=b_{s}/2 \in \mathbb{Z}_{\ge 0}$. 
In case $\Lambda_{j} \in \mathbb{Z}_{\ge 0} $, these parameters are related to 
a $[0,s]$-hook partition $\lambda =(\lambda_{1},\lambda_{2}, \dots )$, $\lambda_{1} \ge \lambda_{2} \ge \dots \ge 0$, 
$\lambda_{1} \le s$: 
\begin{align}
\Lambda_{j}=\lambda_{j}^{\prime} \quad \text{for} \quad j \in \{1,2,\dots , s \}.
\label{YW-B0}
\end{align}
We denote by $V_{\lambda}$ the module $V(\Lambda)$ 
defined in \eqref{YW-B0}. 
The $[0,s]$-hook partition describes a Young 
diagram in the $[0,s]$-hook. This is embedded into the $[0,2s+1]$-hook of $\mathfrak{gl}(0|2s+1)$ (see Figure \ref{MN-hookB0}). 
%%%%%%%%%%
%%%%%%%%%%%%
\begin{figure}
\centering
\begin{tikzpicture}[x=1.6pt,y=1.6pt]
%,line width=1.8pt]
\draw[-to,line width=1.5pt]  (-20,0) -- (220,0);
%\draw[-to,dashed, line width=1.5pt] (80,-60) -- (230,-60);
%\draw[-to,line width=1.5pt] (180,-120) -- (230,-120);
\draw[-to,line width=1.5pt] (0,0) -- (0,-160);
\draw[-to,dashed,line width=1.5pt] (80,0) -- (80,-160);
\draw[-to,line width=1.5pt] (180,0) -- (180,-160);
%
%\draw[dashed]  (0,-60) -- (80,-60);
%\draw[dashed]  (0,-120) -- (180,-120);
\draw[dashed]  (80,0) -- (80,-60);
\draw[dashed]  (180,0) -- (180,-120);
\draw[line width=0.5pt] (80,0) -- (80,-80);
\draw[line width=0.5pt] (60,-80) -- (80,-80);
\draw[line width=0.5pt] (60,0) -- (60,-90);
\draw[line width=0.5pt] (40,-90) -- (60,-90);
\draw[line width=0.5pt] (40,0) -- (40,-110);
\draw[line width=0.5pt] (20,-110) -- (40,-110);
\draw[line width=0.5pt] (20,0) -- (20,-140);
\draw[line width=0.5pt] (0,-140) -- (20,-140);
%\draw[-to,line width=1pt] (32) -- node[midway,left]{$F_{1}^{[2]}$} (22); 
%\draw[dashed] (60,0) -- (140,0);
%\draw[double,double distance=3pt] (200,0) -- (240-6,0);
%
%\draw (240,0) circle (6);
%
\draw (0,0) node[above] {$0$};
\draw (80,0) node[above] {$s$};
\draw (180,0) node[above] {$2s+1$};
\draw (80,7) node[above] {$\lambda_{1}$};
\draw (0,-140) node[left] {$\lambda_{1}^{\prime}$};
\draw (10,-130) node {$ \Lambda_{1}$};
\draw (30,-100) node {$ \Lambda_{2}$};
\draw (50,-80) node {$ \vdots $};
\draw (70,-70) node {$ \Lambda_{s}$};
\draw (0,-160) node[below] {$a$};
\draw (220,0) node[right] {$m$};
\end{tikzpicture} 
\caption{$[0,s]$-hook in $[0,2s+1]$-hook: the Young diagram $\lambda $ is related to the highest weight \eqref{HW-B0} by \eqref{YW-B0}.}
\label{MN-hookB0}
\end{figure}
%%%%%%%%%%%%%%%%%%%%%%%%%%%%%%%%%%%%%%%%%%%%%

%%%%%%%%%%%%%%%%%%%%%%%%%%%%%%%%%%%%%%%%%%%%%%%

\paragraph{Type C and D}
($r,s \in \mathbb{Z}_{\ge 0}$, $r+s \ge 1$)  

The distinguished simple root system of $D(r|s)=\mathfrak{osp}(2r|2s)$, $r>1$ 
is given by
\begin{align}
\begin{split}
\alpha_{i}&=\delta_{i}-\delta_{i+1} \quad \text{for} \quad i \in \{1,2,\dots , s-1\},
\\
\alpha_{s}&=\delta_{s}-\varepsilon_{1},
\\
\alpha_{i+s}&=\varepsilon_{i}-\varepsilon_{i+1} \quad \text{for} \quad i \in \{1,2,\dots , r-1\},
\\
\alpha_{r+s}&=\varepsilon_{r-1}+\varepsilon_{r},
\end{split}
\label{rootDd}
\end{align}
and the corresponding Dynkin diagram has the form

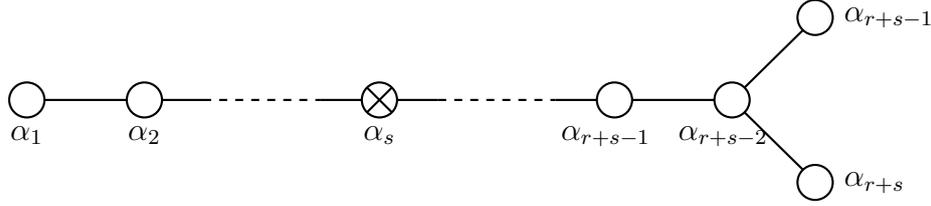
\begin{tikzpicture}[x=1.1pt,y=1.1pt,line width=0.8pt]
\draw  (0,0) circle (6);
\draw (6,0) -- (34,0);
\draw  (40,0) circle (6);
\draw (46,0) -- (60,0);
\draw[dashed] (60,0) -- (100,0);
\draw (100,0) -- (114,0);
\draw  (120,0) circle (6);
\draw (120-4.24,-4.24) -- (120+4.24,4.24);
\draw (120-4.24,4.24) -- (120+4.24,-4.24);
\draw (126,0) -- (140,0);
\draw[dashed] (140,0) -- (180,0);
\draw (180,0) -- (194,0);
\draw (200,0) circle (6);
\draw (206,0) -- (234,0); 
\draw (240,0) circle (6);
\draw (244.24,4.24) -- (264.24,24.04);
\draw (268.28,28.28) circle (6);
\draw (244.24,-4.24) -- (264.24,-24.04);
\draw (268.28,-28.28) circle (6);
\draw (0,-5) node[below] {$\alpha_{1}$};
\draw (40,-5) node[below] {$\alpha_{2}$};
\draw (120,-5) node[below] {$\alpha_{s}$};
\draw (197,-5) node[below] {$\alpha_{r+s-1}$};
\draw (237,-5) node[below] {$\alpha_{r+s-2}$};;
\draw (274.28,28.28) node[right] {$\alpha_{r+s-1}$};
\draw (274.28,-28.28) node[right] {$\alpha_{r+s}$};
\end{tikzpicture} 

%%%%
Let $V(\Lambda )$ be the irreducible module of $\mathfrak{osp}(2r|2s)$ with the highest weight
\begin{align}
\Lambda =\sum_{j=1}^{s} \Lambda_{j} \delta_{j} +\sum_{j=1}^{r} \Lambda_{s+j} \varepsilon_{j} ,
\label{HW-D}
\end{align}
where $\Lambda_{j} \in \mathbb{C}$.
The Kac-Dynkin labels $[b_{1},b_{2},\dots , b_{r+s}]$ of  $V(\Lambda )$ are given by
\begin{align}
b_{j}=\Lambda_{j}-\Lambda_{j+1} \quad \text{for} \quad j \ne s,r+s, \qquad
b_{s}=\Lambda_{s}+\Lambda_{s+1}, \quad b_{r+s}=\Lambda_{r+s-1}+\Lambda_{r+s}. \label{KD-D}
\end{align}
$V(\Lambda )$ is finite dimensional if $b_{j} \in \mathbb{Z}_{\ge 0}$ for $j \ne s$, 
$c=b_{s}-b_{s+1}-b_{s+2}-\cdots -b_{r+s-2}-(b_{r+s-1}+b_{r+s})/2 \in \mathbb{Z}_{\ge 0}$, 
$b_{s+c+1}=b_{s+c+2}=\dots =b_{r+s}=0$ if $c < r-1$, and 
$b_{r+s-1} =b_{r+s}$ if $c =r-1$. 
In case
\footnote{There are misprints in the corresponding part of  
[page 9, \cite{T23}]. In the same paper, $\beta_{s}$ in eq. (2.30) 
 and $\mu^{\prime}_{j}-s$ in eq. (2.33) are  
 misprints of $\beta_{s+1}$ and $\mu^{\prime}_{j}-1$, 
 respectively.}
 $\Lambda_{j} \in \mathbb{Z}_{\ge 0} $ 
($1 \le j \le r+s-1$) and 
$\Lambda_{r+s} \in \mathbb{Z} $, these parameters are related to 
an $[r,s]$-hook partition $\lambda =(\lambda_{1},\lambda_{2}, \dots )$, $\lambda_{1} \ge \lambda_{2} \ge \dots \ge 0$, 
$\lambda_{r+1} \le s$: 
\begin{align}
\Lambda_{j}=\lambda_{j}^{\prime} \quad \text{for} \quad j \in \{1,2,\dots , s \}, \quad 
\Lambda_{s+j}=\max \{ \lambda_{j} -s ,0\} 
\quad \text{for} \quad j \in \{1,2,\dots , r \} ,
\label{YW-Dp}
\end{align}
or 
\begin{align}
\Lambda_{j}&=\lambda_{j}^{\prime} \quad \text{for} \quad j \in \{1,2,\dots , s \}, \quad 
\Lambda_{s+j}=\max \{ \lambda_{j} -s ,0\} 
\quad \text{for} \quad j \in \{1,2,\dots , r-1 \}, 
 \nonumber \\
\Lambda_{r+s}&=-\max \{ \lambda_{r} -s ,0\}  .
\label{YW-Dm}
\end{align}
We denote by $V^{+}_{\lambda}$ (resp. $V^{-}_{\lambda}$) 
the module $V(\Lambda)$ defined in \eqref{YW-Dp} (resp. \eqref{YW-Dm}). 
They coincide when $\Lambda_{r+s}=0$, in which case we simply write $V_{\lambda}$.
 The $[r,s]$-hook partition describes a Young  
diagram in the $[r,s]$-hook. This is embedded into the $[2r,2s+2]$-hook of $\mathfrak{gl}(2r|2s+2)$ (see Figure \ref{MN-hookD}). 
%%%%%%%%%%%%
\begin{figure}
\centering
\begin{tikzpicture}[x=1.6pt,y=1.6pt]
%,line width=1.8pt]
\draw[-to,line width=1.5pt]  (-20,0) -- (230,0);
\draw[-to,dashed, line width=1.5pt] (80,-60) -- (230,-60);
\draw[-to,line width=1.5pt] (200,-120) -- (230,-120);
\draw[-to,line width=1.5pt] (0,0) -- (0,-170);
\draw[-to,dashed,line width=1.5pt] (80,-60) -- (80,-170);
\draw[-to,line width=1.5pt] (200,-120) -- (200,-170);
\draw[dashed]  (0,-60) -- (80,-60);
\draw[dashed]  (0,-120) -- (200,-120);
\draw[dashed]  (80,0) -- (80,-60);
\draw[dashed]  (200,0) -- (200,-120);
\draw[line width=0.5pt] (80,-20) -- (220,-20);
\draw[line width=0.5pt] (220,0) -- (220,-20);
\draw[line width=0.5pt] (80,-40) -- (130,-40);
\draw[line width=0.5pt] (130,-20) -- (130,-40);
\draw[line width=0.5pt] (80,-60) -- (120,-60);
\draw[line width=0.5pt] (120,-40) -- (120,-60);
\draw[line width=0.5pt] (80,0) -- (80,-80);
\draw[line width=0.5pt] (60,-80) -- (80,-80);
\draw[line width=0.5pt] (60,0) -- (60,-90);
\draw[line width=0.5pt] (40,-90) -- (60,-90);
\draw[line width=0.5pt] (40,0) -- (40,-110);
\draw[line width=0.5pt] (20,-110) -- (40,-110);
\draw[line width=0.5pt] (20,0) -- (20,-140);
\draw[line width=0.5pt] (0,-140) -- (20,-140);
%\draw[-to,line width=1pt] (32) -- node[midway,left]{$F_{1}^{[2]}$} (22); 
%\draw[dashed] (60,0) -- (140,0);
%\draw[double,double distance=3pt] (200,0) -- (240-6,0);
%
%\draw (240,0) circle (6);
%
\draw (0,0) node[above] {$0$};
\draw (80,0) node[above] {$s$};
\draw (200,0) node[above] {$2s+2$};
\draw (220,0) node[above] {$\lambda_{1}$};
\draw (0,-60) node[left] {$r$};
\draw (0,-120) node[left] {$2r$};
\draw (0,-140) node[left] {$\lambda_{1}^{\prime}$};
\draw (210,-10) node {$\Lambda_{s+1}$};
\draw (120,-30) node {$ \cdots $};
\draw (110,-50) node {$ |\Lambda_{s+r}|$};
\draw (10,-130) node {$ \Lambda_{1}$};
\draw (30,-100) node {$ \Lambda_{2}$};
\draw (50,-80) node {$ \vdots $};
\draw (70,-70) node {$ \Lambda_{s}$};
\draw (0,-170) node[below] {$a$};
\draw (230,0) node[right] {$m$};
\end{tikzpicture} 
\caption{$[r,s]$-hook in $[2r,2s+2]$-hook: the Young diagram $\lambda $ is related to the highest weight \eqref{HW-D} by \eqref{YW-Dp} or  \eqref{YW-Dm}.}
\label{MN-hookD}
\end{figure}
%%%%%%%%%%%%%%%%%%%%%%%%%%%%%%%%%%%%%%%%%%%%%
%%%%%%%%%%%%%%%%%%%%%%%%%%

\subparagraph{Type C for $r=1$, $s \ge 1$:}
The distinguished simple root system of $C(s+1)=\mathfrak{osp}(2|2s)$ 
is given by 
\begin{align}
\alpha_{1}&=\varepsilon_{1}-\delta_{1} , \\
\alpha_{i}&=\delta_{i-1}-\delta_{i} \quad \text{for} \quad i \in \{2,3,\dots , s\},
\\
\alpha_{s+1}&=2\delta_{s},
\end{align}
%%%%%
and the corresponding Dynkin diagram has the form

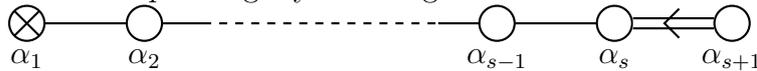
\begin{tikzpicture}[x=1.1pt,y=1.1pt,line width=0.8pt]
\draw  (0,0) circle (6);
\draw (-4.24,-4.24) -- (4.24,4.24);
\draw (-4.24,4.24) -- (4.24,-4.24);
\draw (6,0) -- (34,0);
\draw  (40,0) circle (6);
\draw (46,0) -- (60,0);
\draw[dashed] (60,0) -- (140,0);
\draw (140,0) -- (154,0);
\draw (160,0) circle (6);
\draw (166,0) -- (194,0);
\draw (200,0) circle (6);
\draw[double,double distance=3pt] (206.3,0) -- (233.8,0);
\draw (217+5,-5) -- (217,0) -- (217+5,5);
\draw (240,0) circle (6);
\draw (0,-5) node[below] {$\alpha_{1}$};
\draw (40,-5) node[below] {$\alpha_{2}$};
\draw (160,-5) node[below] {$\alpha_{s-1}$};
\draw (200,-5) node[below] {$\alpha_{s}$};
\draw (240,-5) node[below] {$\alpha_{s+1}$};
\end{tikzpicture} 
%%%%%%%%%

\begin{figure}
\centering
\begin{tikzpicture}[x=1.6pt,y=1.6pt]
%,line width=1.8pt]
\draw[-to,line width=1.5pt]  (-20,0) -- (230,0);
\draw[-to,dashed, line width=1.5pt] (80,-20) -- (230,-20);
\draw[-to,line width=1.5pt] (200,-40) -- (230,-40);
\draw[-to,line width=1.5pt] (0,0) -- (0,-170);
\draw[-to,dashed,line width=1.5pt] (80,-40) -- (80,-170);
\draw[-to,line width=1.5pt] (200,-40) -- (200,-170);
\draw[dashed]  (0,-20) -- (80,-20);
\draw[dashed]  (0,-40) -- (200,-40);
\draw[dashed]  (80,0) -- (80,-60);
\draw[dashed]  (200,0) -- (200,-40);
\draw[line width=0.5pt] (0,-20) -- (220,-20);
\draw[line width=0.5pt] (220,0) -- (220,-20);
\draw[line width=0.5pt] (80,-20) -- (80,-80);
\draw[line width=0.5pt] (60,-80) -- (80,-80);
\draw[line width=0.5pt] (60,-20) -- (60,-90);
\draw[line width=0.5pt] (40,-90) -- (60,-90);
\draw[line width=0.5pt] (40,-20) -- (40,-110);
\draw[line width=0.5pt] (20,-110) -- (40,-110);
\draw[line width=0.5pt] (20,-20) -- (20,-140);
\draw[line width=0.5pt] (0,-140) -- (20,-140);
%\draw[-to,line width=1pt] (32) -- node[midway,left]{$F_{1}^{[2]}$} (22); 
%\draw[dashed] (60,0) -- (140,0);
%\draw[double,double distance=3pt] (200,0) -- (240-6,0);
%
%\draw (240,0) circle (6);
%
\draw (0,0) node[above] {$0$};
\draw (80,0) node[above] {$s$};
\draw (200,0) node[above] {$2s+2$};
\draw (220,0) node[above] {$\lambda_{1}$};
\draw (0,-20) node[left] {$1$};
\draw (0,-40) node[left] {$2$};
\draw (0,-140) node[left] {$\lambda_{1}^{\prime}$};
\draw (210,-10) node {$\Lambda_{1}$};
\draw (10,-130) node {$ \Lambda_{2}$};
\draw (30,-100) node {$ \Lambda_{3}$};
\draw (50,-80) node {$ \vdots $};
\draw (70,-70) node {$ \Lambda_{s+1}$};
\draw (0,-170) node[below] {$a$};
\draw (230,0) node[right] {$m$};
\end{tikzpicture} 
\caption{$[1,s]$-hook in $[2,2s+2]$-hook: the Young diagram $\lambda $ is related to the highest weight \eqref{HW-Cp} by \eqref{YW-Cp}.}
\label{MN-hookCp}
\end{figure}
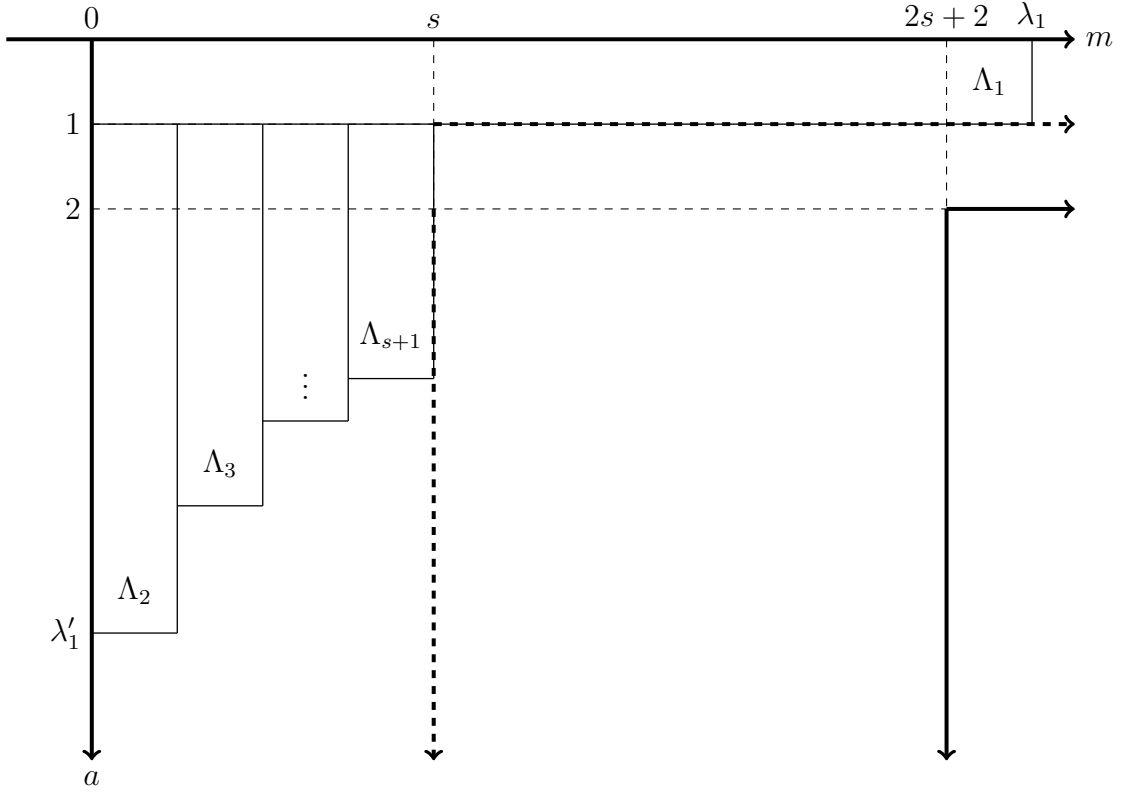

%%%%
Let $V(\Lambda )$ be the irreducible module of $\mathfrak{osp}(2|2s)$ with the highest weight
\begin{align}
\Lambda =\Lambda_{1} \varepsilon_{1} +\sum_{j=1}^{s} \Lambda_{1+j} \delta_{j} ,
\label{HW-Cp}
\end{align}
where $\Lambda_{j} \in \mathbb{C}$.
The Kac-Dynkin labels $[b_{1},b_{2},\dots , b_{s+1}]$ of  $V(\Lambda )$ are given by
\begin{align}
b_{1}=\Lambda_{1}+\Lambda_{2}, \qquad 
b_{j}=\Lambda_{j}-\Lambda_{j+1} \quad \text{for} \quad j \ne 1,s+1, \qquad
b_{s+1}=\Lambda_{s+1}. \label{KD-C}
\end{align}
$V(\Lambda )$ is finite dimensional if $b_{j} \in \mathbb{Z}_{\ge 0}$ for $j \ne 1$.  
In case $\Lambda_{j} \in \mathbb{Z}_{\ge 0} $, these parameters are related to 
a $[1,s]$-hook partition $\lambda =(\lambda_{1},\lambda_{2}, \dots )$, $\lambda_{1} \ge \lambda_{2} \ge \dots \ge 0$, 
$\lambda_{2} \le s$: 
\begin{align}
\Lambda_{1}= \lambda_{1}, \qquad 
\Lambda_{j+1}= \max \{ \lambda_{j}^{\prime}-1 ,0\},  
\quad \text{for} \quad j \in \{1,2,\dots , s \} .
\label{YW-Cp}
\end{align}
We denote by $V_{\lambda}$ the module $V(\Lambda)$ 
defined in \eqref{YW-Cp}. 
 The $[1,s]$-hook partition describes a Young 
diagram in the $[1,s]$-hook. This is embedded into the $[2,2s+2]$-hook of $\mathfrak{gl}(2|2s+2)$ (see Figure \ref{MN-hookCp}). 
%%%%%%%%%%%%
%%%%%%%%%%%%%%%%%%%%%%%%%%%%%%%%%%%%%%%%%%%%%
%%%%%
\section{Decomposition of the supersymmetric Schur function}
Let $X=\{x_{1},x_{2},\dots\}$, $Y=\{y_{1},y_{2},\dots\}$, and $T=\{t_{1},t_{2},\dots\}$ be sets of indeterminates, and let $t$ be a single indeterminate. In this section, we review identities for (super)symmetric functions in these variables and reformulate them so that they fit our purposes.
A standard reference on symmetric functions is \cite{Mac98}.
References \cite{BB81,Remmel84,CK87,BSR96,Okada06} are also relevant to our discussion.
In particular, we will frequently refer to \cite{Okada06}, where detailed proofs of several identities in the special case $Y=\emptyset$ are given.

The complete (super)symmetric function is defined by the following generating function:
\begin{align}
\sum_{j=0}^{\infty} h_{j}(X|Y)t^{j}=\frac{\prod_{j}(1-y_{j}t)}{\prod_{j}(1-x_{j}t)} ,
\label{genh}
\end{align}
where $h_{0}(X|Y)=1$, and $h_{j}(X|Y)=0$ if $j<0$. 

For any non-empty partition $\lambda \in \mathcal{P}$, we define 
Jacobi-Trudi-type determinants:
%%%%%%%%%%
\footnote{
%%%%%%%%%%%%%%%
One can also rewrite \eqref{JT-B} and \eqref{JT-C} as 
\begin{align}
S_{[ \lambda ] }  (X|Y)
&=\det (
(H_{\lambda_{i}-i+1}(X|Y))_{1 \le i \le \lambda_{1}^{\prime}} \
 (H_{\lambda_{i}-i+j}(X|Y) +H_{\lambda_{i}-i-j+2}(X|Y))_{1 \le i \le \lambda_{1}^{\prime} \atop 2 \le j \le \lambda_{1}^{\prime} } )
 \label{JT-b}
\\
&=\frac{1}{2}\det_{1 \le i,j \le \lambda_{1}^{\prime}} (
 H_{\lambda_{i}-i+j}(X|Y) +H_{\lambda_{i}-i-j+2}(X|Y)  ),
\\
S_{\langle \lambda \rangle }  (X|Y)&=\det (
(h_{\lambda_{i}-i+1}(X|Y))_{1 \le i \le \lambda_{1}^{\prime}} \
 (h_{\lambda_{i}-i+j}(X|Y) +h_{\lambda_{i}-i-j+2}(X|Y))_{1 \le i \le \lambda_{1}^{\prime} \atop 2 \le j \le \lambda_{1}^{\prime} } ),
\label{JT-c}
\end{align}
 where $H_{m}=h_{m}(X|Y)-h_{m-2}(X|Y)$. 
\eqref{JT-b} and \eqref{JT-c} are the determinants of block matrices consisting of 
a $\lambda_{1}^{\prime} \times 1$ matrix and a $\lambda_{1}^{\prime} \times (\lambda_{1}^{\prime}-1)$ matrix. 
%%%%%%%%%%%%%%
}
%%%%%%%% 
\begin{align}
S_{\lambda}(X|Y)&=\det_{1 \le i,j \le l(\lambda)} (h_{\lambda_{i}-i+j}(X|Y)), \label{JT-A} \\
S_{[\lambda]}(X|Y)&=\det_{1 \le i,j \le l(\lambda)}
 (h_{\lambda_{i}-i+j}(X|Y)-h_{\lambda_{i}-i-j}(X|Y)), \label{JT-B} \\
S_{\langle \lambda \rangle}(X|Y)&=
\frac{1}{2}\det_{1 \le i,j \le l(\lambda)}
 (h_{\lambda_{i}-i+j}(X|Y)+h_{\lambda_{i}-i-j+2}(X|Y)), \label{JT-C} 
\end{align}
and set $S_{\emptyset}(X|Y)=S_{[\emptyset]}(X|Y)=
S_{\langle \emptyset \rangle}(X|Y)=1$.
In particular, \eqref{JT-A}, which is known as the supersymmetric
Schur function, reduces to the ordinary Schur function
$S_{\lambda}(X)=S_{\lambda}(X| \emptyset)$ when $Y=\emptyset$.
Note that these functions are not changed by adding the same element $\eta$ to the sets $X$ and $Y$ at the same time, since the cancellation occurs in the right-hand side of \eqref{genh}:
\begin{align}
S_{\lambda}(X \sqcup \{\eta \} |Y \sqcup \{\eta \})&=S_{\lambda}(X|Y), 
\label{SP-A} \\
S_{[\lambda]}(X\sqcup \{\eta \}|Y\sqcup \{\eta \})&=S_{[\lambda]}(X|Y), \label{SP-B} \\
S_{\langle \lambda \rangle}(X\sqcup \{\eta \}|Y\sqcup \{\eta \})&=S_{\langle \lambda \rangle}(X|Y). \label{SP-C} 
\end{align}
For $\mu , \nu \in \mathcal{P}$, the Schur function satisfies 
\begin{align}
S_{\mu}(T) S_{\nu}(T)=
\sum_{\lambda \in \mathcal{P}}
\mathrm{LR}^{\lambda}_{\nu , \mu} S_{\lambda}(T),
\label{LR}
\end{align}
where $\mathrm{LR}^{\lambda}_{\nu , \mu} $ is the Littlewood-Richardson coefficient. 
We are interested in the following properties of the 
Littlewood-Richardson coefficient
 [Proposition 9.21, Corollary 9.23, Exercise 9.14 in page 165, \cite{Okada06}]:
  for $\lambda, \mu, \nu \in {\mathcal P}$, 
\begin{align}
\mathrm{LR}^{\lambda}_{\mu , \nu}&=\mathrm{LR}^{\lambda}_{\nu , \mu}
\label{LR1}
\\
\mathrm{LR}^{\lambda}_{\emptyset , \nu}&=
\mathrm{LR}^{\lambda}_{\nu , \emptyset}=\delta_{\lambda, \nu},
\\
\mathrm{LR}^{\lambda}_{\mu , \nu}&=0 \quad \text{unless} \quad 
|\lambda|=|\mu|+|\nu | ,
\\
\mathrm{LR}^{\mu + \nu}_{\mu , \nu}&=1,
\\
\mathrm{LR}^{\lambda^{\prime}}_{\mu^{\prime} , \nu^{\prime}}&= 
\mathrm{LR}^{\lambda}_{\mu , \nu} ,
\label{LRtran}
\end{align}
where $\mu + \nu =(\mu_{1}+\nu_{1},\mu_{2}+\nu_{2}, \dots )$, 
and in particular for $\mu,\nu \subset (m^{a})$  ($m,a \in {\mathbb Z}_{\ge 1}$),
\begin{align}
\mathrm{LR}^{(m^{a})}_{\mu , \nu}&=
\begin{cases}
1 & \text{if} \quad \mu_{i}+\nu_{a+1-i}=m  \quad \text{for all}
 \quad 1 \le i \le m \\
0 & \text{otherwise}.
\end{cases}
\label{LR5}
\end{align}
We will use the Schur's formula
\begin{align}
\sum_{\lambda \in \mathcal{P}} S_{\lambda}(T)
=\frac{1}{\prod_{i}(1-t_{i}) \prod_{i<j }
(1-t_{i}t_{j})},
\label{sum-S}
\end{align}
and the Littlewood's formulas [Theorem 9.28, \cite{Okada06}]
\begin{align}
\sum_{\lambda \in {\mathcal P}^{+}} S_{\lambda}(T)
&=\frac{1}{\prod_{i\le j }
(1-t_{i}t_{j})},
\label{sum-L1}
\\
\sum_{\lambda \in {\mathcal P}^{-}} S_{\lambda}(T)
&=\frac{1}{\prod_{ i<j }
(1-t_{i}t_{j})} .
\label{sum-L2}
\end{align}
We will also use the homogeneity property of 
the (supersymmetric) Schur function:
\begin{align}
S_{\lambda}(\xi  X)=\xi^{|\lambda|}S_{\lambda}(X),
\qquad 
S_{\lambda}(\xi  X|\xi  Y)=\xi^{|\lambda|}S_{\lambda}(X|Y),
\label{signS}
\end{align}
where $\xi  X=\{\xi x_{1},\xi x_{2},\dots \}$ and 
$\xi  Y=\{\xi y_{1},\xi y_{2},\dots \}$.
%%%%%%%%%%%
%From now on, we shall employ a simplified notation in which symbols such as $\sqcup$ and $\{\}$ are suppressed in the arguments of \eqref{genh}-\eqref{JT-B}. For example, will use 
%$(X \sqcup Z|Y \sqcup \{1,-1\})=(X,Z|Y,1,-1)$ for sets $X,Y,Z$ and $\{1,-1\}$. 

\begin{proposition}\label{pro-gen}
Let $\xi = \pm 1$ and $\lambda \in {\mathcal P}$.
The following relations are valid. 
\begin{align}
S_{\lambda}(X|Y)&=\sum_{\mu \in {\mathcal P}}
\sum_{\kappa \in {\mathcal P}^{+} } \mathrm{LR}^{\lambda}_{\kappa , \mu} 
S_{[\mu]}(X|Y)
\label{dc-1a}
\\
&=\sum_{\mu \in {\mathcal P}}
\sum_{\kappa \in {\mathcal P}^{-} }
 \mathrm{LR}^{\lambda}_{\kappa , \mu} 
S_{\langle \mu \rangle}(X |Y),
\label{dc-1b}
\end{align}
%%%%%%%%%%%%
\begin{align}
S_{\lambda}(X|Y \sqcup \{\xi \})&=\sum_{\mu \in {\mathcal P}}
\sum_{\kappa \in {\mathcal P}^{-} } \mathrm{LR}^{\lambda}_{\kappa , \mu} 
S_{[\mu]}(X \sqcup \{-\xi \} |Y)
\label{dc-2a}
\\
&=\sum_{\mu,\nu \in {\mathcal P}}
(-\xi)^{|\nu|} \mathrm{LR}^{\lambda}_{\nu , \mu} 
S_{[\mu]}(X |Y),
\label{dc-2b}
\end{align}
%%%%%%%%%%%%
\begin{align}
S_{\lambda}(X \sqcup \{\xi \}|Y)&=\sum_{\mu \in {\mathcal P}}
\sum_{\kappa \in {\mathcal P}^{+} } \mathrm{LR}^{\lambda}_{\kappa , \mu} 
S_{\langle \mu \rangle}(X |Y\sqcup \{-\xi \})
\label{dc-3a}
\\
&=\sum_{\mu,\nu \in {\mathcal P}}
\xi^{|\nu|} \mathrm{LR}^{\lambda}_{\nu , \mu} 
S_{\langle \mu \rangle}(X |Y),
\label{dc-3b}
\end{align}
%%%%%%%%%%%%
\begin{align}
S_{\lambda}(X|Y\sqcup \{1,-1\})&=\sum_{\mu \in {\mathcal P}}
\sum_{\kappa \in {\mathcal P}^{-} } \mathrm{LR}^{\lambda}_{\kappa , \mu} 
S_{[\mu ]}(X |Y),
\label{dc-4}
\end{align}
\begin{align}
S_{\lambda}(X\sqcup \{1,-1\}|Y)&=\sum_{\mu \in {\mathcal P}}
\sum_{\kappa \in {\mathcal P}^{+} } \mathrm{LR}^{\lambda}_{\kappa , \mu} 
S_{\langle \mu \rangle}(X |Y).
\label{dc-4du}
\end{align}
\end{proposition}
%%%%%%%%%%%%
%\begin{align}
%S_{\lambda}(X \sqcup \{\xi \} |Y \sqcup \{1,-1\})&=
%\sum_{\mu \in {\mathcal P}}
%\sum_{\kappa \in {\mathcal P}^{+} } 
%\mathrm{LR}^{\lambda}_{\kappa , \mu} 
%S_{[\mu]}(X |Y \sqcup \{-\xi \} ),
%\label{dc-5a}
%\\
%%
%&=\sum_{\mu \in {\mathcal P}}
%\sum_{\kappa \in {\mathcal P}^{-} }
% \mathrm{LR}^{\lambda}_{\kappa , \mu} 
%S_{\langle \mu \rangle}(X |Y \sqcup \{-\xi \}),
%\label{dc-5b}
%\\
%%
%&=\sum_{\mu ,\nu \in {\mathcal P}}
%\xi^{|\nu|}
% \mathrm{LR}^{\lambda}_{\nu , \mu} 
%S_{[\mu]}(X |Y ),
%\label{dc-5c}
%\end{align}
%%%%%%%%%%%%%%%%%%%
In order to prove \eqref{dc-1a}-\eqref{dc-4du}, we need the 
following Cauchy-type identities: 
\begin{lemma}\label{Cauchy-lem}
The following relations are valid.
\begin{align}
\sum_{\lambda \in {\mathcal P}}
S_{\lambda}(X|Y)S_{\lambda }(T)
&=\prod_{j}
\left(
\frac{\prod_{i} (1-t_{j}y_{i})}{\prod_{i} (1-t_{j}x_{i})}
\right) ,
\label{co-A}
\\
\sum_{\lambda \in {\mathcal P}}
S_{[\lambda]}(X|Y)S_{\lambda }(T)
&=\prod_{j }
\left(
\frac{\prod_{i} (1-t_{j}y_{i})}{\prod_{i} (1-t_{j}x_{i})}
\right)
 \prod_{i \le j}(1-t_{i}t_{j}),
\label{co-B}
\\
\sum_{\lambda \in {\mathcal P}}
S_{\langle \lambda \rangle}(X|Y)S_{\lambda }(T)
&=\prod_{j}
\left(
\frac{\prod_{i} (1-t_{j}y_{i})}{\prod_{i} (1-t_{j}x_{i})}
\right)
 \prod_{i < j}(1-t_{i}t_{j}).
\label{co-C}
\end{align}
\end{lemma}
These formulas for finite sets $X$ and $Y$ are generating functions 
of supercharacters of tensor representations 
of the general linear \eqref{co-A} 
or the orthosymplectic (\eqref{co-B}, \eqref{co-C}) 
superalgebras, and 
in particular, 
\eqref{co-A} corresponds to 
[a special case of eq.\ (1.7) in \cite{Remmel84}; 
\eqref{co-B} corresponds to [eq.\ (3.15) in \cite{CK87}];  
\eqref{co-C} corresponds to [eq.\ (c), Theorem 4.24 in \cite{BSR96}]. 
Moreover, an analogous Cauchy-type identity holds for 
$S_{\langle \lambda^{\prime} \rangle}(X|Y)$ 
[eq.\ (d), Theorem 4.24 in \cite{BSR96}]: 
\begin{align}
 \sum_{\lambda \in {\mathcal P}}
S_{\langle \lambda^{\prime} \rangle}(X|Y)S_{\lambda }(T)
&=\prod_{j}
\left(
\frac{\prod_{i} (1+t_{j}x_{i})}{\prod_{i} (1+t_{j}y_{i})}
\right)
 \prod_{i \le j}(1-t_{i}t_{j}).
\label{co-Cdual}
\end{align}
By comparing \eqref{co-B} and \eqref{co-Cdual}, 
one finds
\begin{align}
S_{\langle \lambda^{\prime} \rangle}(X|Y)&=
S_{[ \lambda ]}(-Y|-X)=(-1)^{|\lambda|}S_{[ \lambda ]}(Y|X). 
\label{BCD-dual}
\end{align}
A similar identity is also known [eq.\ (1.6), \cite{Remmel84}]:
\begin{align}
S_{\lambda^{\prime}}(X|Y)&=
S_{ \lambda }(-Y|-X)=(-1)^{|\lambda|}S_{ \lambda }(Y|X).
 \label{A-dual}
\end{align}
We present a proof of \eqref{co-B}, which parallels the case 
$Y=0$ as described in [Proposition 12.3, \cite{Okada06}]. 
Here the sets $X,Y$ and $T$ are not necessarily assumed to be finite, in a manner that extends naturally to our setting.
The proofs of \eqref{co-A}, \eqref{co-C} and \eqref{co-Cdual} are similar to that of \eqref{co-B} (cf. \cite{BSR96}).

\begin{proof}[Proof of \eqref{co-B} in Lemma \ref{Cauchy-lem}]
It suffices to show that the following relation holds 
for any $m \in {\mathbb Z}_{\ge 1}$:
\begin{align}
\sum_{\lambda \in {\mathcal P}}
S_{[\lambda]}(X|Y)S_{\lambda }(T^{(m)})
&=\prod_{j=1}^{m}
\left(
\frac{\prod_{i} (1-t_{j}y_{i})}{\prod_{i} (1-t_{j}x_{i})}
\right)
 \prod_{1 \le i \le j \le m}(1-t_{i}t_{j}),
\label{co-Bm}
\end{align}
where $T^{(m)}=\{t_{1},t_{2},\dots, t_{m}\}$.
Substituting the Schur function
\begin{align}
S_{\lambda }(T^{(m)})=
\frac{\det_{1 \le i,j \le m} (t_{i}^{\lambda_{j}+m-j})}
{\det_{1 \le i,j \le m} (t_{i}^{m-j})}
\label{SchDet}
\end{align}
into the left hand side of \eqref{co-Bm}, and multiplying 
the Vandermonde determinant 
\begin{align}
\det_{1 \le i,j \le m} (t_{i}^{m-j})
=\prod_{1 \le i < j \le m}(t_{i}-t_{j}) 
\label{Van}
\end{align}
on both sides, we obtain
\begin{multline}
\sum_{\lambda \in {\mathcal P}}
S_{[\lambda]}(X|Y)\det_{1 \le i,j \le m} (t_{i}^{\lambda_{j}+m-j})
=
\\
=\prod_{j=1}^{m}
\left(
\frac{\prod_{i} (1-t_{j}y_{i})}{\prod_{i} (1-t_{j}x_{i})}
\right)
 \prod_{i=1}^{m}(1-t_{i}^{2})
\prod_{1 \le i < j \le m}^{m}(t_{i}-t_{j})(1-t_{i}t_{j}).
\label{co-Bm2}
\end{multline}
In order to prove this, we need the following identity
\begin{align}
\det_{1 \le i,j \le m} (t_{i}^{m-j}-t_{i}^{m+j})
=\prod_{i=1}^{m}(1-t_{i}^{2})
\prod_{1 \le i < j \le m}^{m}(t_{i}-t_{j})(1-t_{i}t_{j}).
\label{mmdet}
\end{align}
By setting $t_{i} \to t_{i}+t_{i}^{-1}$ in \eqref{Van}, we obtain 
\begin{align}
\det_{1 \le i,j \le m} ((t_{i}+t_{i}^{-1})^{m-j})
=(-1)^{\frac{m(m-1)}{2}}\prod_{i=1}^{m}t_{i}^{1-m}
\prod_{1 \le i < j \le m}^{m}(t_{i}-t_{j})(1-t_{i}t_{j}).
\label{mmdet0}
\end{align}
The $(i,j)$-matrix element of the left hand side of \eqref{mmdet0} 
can be expanded as
\begin{multline}
(t_{i}+t_{i}^{-1})^{m-j}
= \sum_{k=0}^{m-j}\binom{m-j}{k}t_{i}^{2k-m+j}
\\
=
\begin{cases}
1 & \text{if} \quad m-j=0
\\
t_{i}^{m-j}+t_{i}^{-m+j}+
\sum_{k=1}^{\frac{m-j}{2}-1}\binom{m-j}{k}
(t_{i}^{-2k+m-j}+t_{i}^{2k-m+j})+\binom{m-j}{\frac{m-j}{2}} & \text{if} \quad m-j \in 2\mathbb{Z}_{\ge 1}
\\
t_{i}^{m-j}+t_{i}^{-m+j}+
\sum_{k=1}^{\frac{m-j-1}{2}}\binom{m-j}{k}
(t_{i}^{-2k+m-j}+t_{i}^{2k-m+j}) &
 \! \! \! \! 
 \text{if} \quad m-j \in 2\mathbb{Z}_{\ge 0}+1
\end{cases}
\label{expan}
\end{multline}
Thus, in the left hand side of \eqref{mmdet0}, 
by repeatedly performing the operation of subtracting 
$\binom{m-j}{k}$ times the $(j+2k)$-th column from the 
 $j$-th column of the matrix, for $k=1,2,\dots , (m-j)/2$ 
 if $m-j \in 2\mathbb{Z}_{\ge 1}$, and 
 for $k=1,2,\dots , (m-j-1)/2$ 
 if $m-j \in 2\mathbb{Z}_{\ge 1}+1$, 
 starting from $j=m-2$ down to $j=1$, 
we arrive at the following relation: 
\begin{align}
\det_{1 \le i,j \le m} ((t_{i}+t_{i}^{-1})^{m-j})
&=\frac{1}{2}\det_{1 \le i,j \le m} (t_{i}^{m-j}+t_{i}^{-m+j}).
\label{mmdet1}
\\
&=
\frac{(-1)^{\frac{m(m-1)}{2}}}{2}\det_{1 \le i,j \le m} (t_{i}^{j-1}+t_{i}^{-j+1}).
\label{mmdet2}
\end{align}
Note that the prefactor $1/2$ in the right hand side of 
\eqref{mmdet1} is necessary since 
$(t_{i}^{m-j}+t_{i}^{-m+j})$ becomes $2$ at $j=m$ while \eqref{expan} becomes $1$. 
Multiplying $(-1)^{\frac{m(m-1)}{2}}\prod_{i=1}^{m}t_{i}^{m-1}
(1-t_{i}^{2})$ on both sides of \eqref{mmdet1}, we obtain
\begin{multline}
(-1)^{\frac{m(m-1)}{2}}\prod_{i=1}^{m}t_{i}^{m-1}
(1-t_{i}^{2})
\det_{1 \le i,j \le m} ((t_{i}+t_{i}^{-1})^{m-j})
\\
=
\frac{1}{2}\det_{1 \le i,j \le m} (t_{i}^{m+j-2}-t_{i}^{m+j}+
t_{i}^{m-j}-t_{i}^{m-j+2}).
\label{mmdet3}
\end{multline}
Thus, in the right hand side of \eqref{mmdet3}, 
after multiplying the first column of the matrix by the prefactor 
$1/2$,
we perform a sequence of column operations in strictly descending order of the column index $j$, starting from $j=m$ down to $j=3$.
Specifically, for each such $j$, we add the $(j-2k)$-th column 
to the $j$-th column, where $k=1,2,\dots ,(j-1)/2$ 
 if $j \in 2\mathbb{Z}_{\ge 1}+1$, and 
 for $k=1,2,\dots ,j/2-1$ if $j \in 2\mathbb{Z}_{\ge 2}$.
Through this process, we find that \eqref{mmdet3} is equivalent to 
$\det_{1 \le i,j \le m} (t_{i}^{m-j}-t_{i}^{m+j})$, 
and thus \eqref{mmdet0} is equivalent to \eqref{mmdet}.

Let us modify the right hand side of \eqref{co-Bm2} as
\begin{multline}
\text{r.h.s. of \eqref{co-Bm2}}=
\prod_{j=1}^{m}
\left(
\frac{\prod_{i} (1-t_{j}y_{i})}{\prod_{i} (1-t_{j}x_{i})}
\right)
\det_{1 \le i,j \le m} (t_{i}^{m-j}-t_{i}^{m+j})
\quad \text{[by \eqref{mmdet}]}
\\
=\prod_{j=1}^{m}
\left(
\sum_{k_{j}=0}^{\infty} h_{k_{j}}(X|Y)t^{k_{j}}_{j}
\right)
\det_{1 \le i,j \le m} (t_{i}^{m-j}-t_{i}^{m+j})
 \quad \text{[by \eqref{genh}]}
\\
=
\left(
\sum_{k_{1}=0}^{\infty}
\dots 
\sum_{k_{m}=0}^{\infty}
\prod_{i=1}^{m}
 h_{k_{i}}(X|Y)t^{k_{i}}_{i}
\right)
\sum_{\sigma \in \mathcal{S}_{m}}
\mathrm{sgn}(\sigma)
\prod_{i=1}^{m}
(t_{i}^{m-\sigma(i)}-t_{i}^{m+\sigma(i)})
\\
=
\sum_{k_{1}=0}^{\infty}
\dots 
\sum_{k_{m}=0}^{\infty}
\sum_{\sigma \in \mathcal{S}_{m}}
\sum_{\xi_{1} \in \{-1,1\}}
\dots
\sum_{\xi_{m} \in \{-1,1\}}
\mathrm{sgn}(\sigma)
\prod_{i=1}^{m}
 h_{k_{i}}(X|Y)\xi_{i} t^{k_{i}+m-\xi_{i}\sigma(i)}_{i},
\end{multline}
where $\mathcal{S}_{m}$ denotes the symmetric group on $\{1,2,\dots ,m \}$.
%%%
By setting $\eta_{i}=k_{i}+m-\xi_{i}\sigma(i)$, we obtain
\begin{multline}
=
\sum_{\eta_{1}=0}^{\infty}
\dots 
\sum_{\eta_{m}=0}^{\infty}
\sum_{\sigma \in \mathcal{S}_{m}}
\sum_{\xi_{1} \in \{-1,1\}}
\dots
\sum_{\xi_{m} \in \{-1,1\}}
\mathrm{sgn}(\sigma)
\prod_{i=1}^{m}
 h_{\eta_{i}-m+\xi_{i}\sigma(i)}(X|Y)\xi_{i} t^{\eta_{i}}_{i}
 \\
=
\sum_{\eta_{1}=0}^{\infty}
\dots 
\sum_{\eta_{m}=0}^{\infty}
\sum_{\sigma \in \mathcal{S}_{m}}
\mathrm{sgn}(\sigma)
\prod_{i=1}^{m}
 (h_{\eta_{i}-m+\sigma(i)}(X|Y)-
 h_{\eta_{i}-m-\sigma(i)}(X|Y)) t^{\eta_{i}}_{i}
 \\
=
\sum_{\eta_{1}=0}^{\infty}
\dots 
\sum_{\eta_{m}=0}^{\infty}
\det_{1 \le i,j \le m} (h_{\eta_{i}-m+j}(X|Y)-
 h_{\eta_{i}-m-j}(X|Y)) \prod_{i=1}^{m} t^{\eta_{i}}_{i} .
\end{multline}
If any of the elements of $\{\eta_{1},\eta_{2},\dots ,\eta_{m}\}$ are equal to each other, the determinant above becomes zero, so by rearranging the rows, it takes the following form.
\begin{multline}
=
\sum_{\eta_{1}>\eta_{2} >\dots >\eta_{m} \ge 0}
\sum_{\sigma \in \mathcal{S}_{m}}
\mathrm{sgn}(\sigma)
\det_{1 \le i,j \le m} (h_{\eta_{i}-m+j}(X|Y)-
 h_{\eta_{i}-m-j}(X|Y)) \prod_{i=1}^{m} t^{\eta_{\sigma(i)}}_{i}
\\
=
\sum_{\eta_{1}>\eta_{2} >\dots >\eta_{m} \ge 0}
\det_{1 \le i,j \le m} (h_{\eta_{i}-m+j}(X|Y)-
 h_{\eta_{i}-m-j}(X|Y)) 
 \det_{1 \le i,j \le m}( t^{\eta_{j}}_{i}).
\end{multline}
%%%
By setting 
$\eta_{i}=\lambda_{i}+m-i$, we obtain 
\begin{multline}
=
\sum_{\lambda_{1} \ge \lambda_{2} \ge
 \dots \ge \lambda_{m} \ge 0}
\det_{1 \le i,j \le m} (h_{\lambda_{i}-i+j}(X|Y)-
 h_{\lambda_{i}-i-j}(X|Y)) 
 \det_{1 \le i,j \le m}( t^{\lambda_{j}+m-j}_{i})
 \\
 =\text{[left hand side of \eqref{co-Bm2}]}.
\end{multline}
This proves \eqref{co-Bm}, and thus \eqref{co-B}. 
\end{proof}
%%%%%%%%%%%%%%%%%%%%%%%%%
\begin{proof}[Proof of Proposition \ref{pro-gen}]
Let us set $(X,Y)\to (X  ,Y \sqcup \{1,-1 \})$ 
in \eqref{co-A}, and modify this as follows:
\begin{multline}
\sum_{\lambda \in {\mathcal P}}
S_{\lambda}(X |Y \sqcup 
 \{1,-1\}) S_{\lambda }(T)
=\prod_{j}
\left(
\frac{(1-t_{j})(1+t_{j}) \prod_{i}(1-t_{j}y_{i})}
{\prod_{i} (1-t_{j}x_{i})}
\right)
 \\ 
=\prod_{j }
\left(
\frac{ \prod_{i}(1-t_{j}y_{i})}{\prod_{i} (1-t_{j}x_{i})}
\right)\prod_{i \le j}(1-t_{i}t_{j}) 
 \times \frac{1}{\prod_{i < j}(1-t_{i}t_{j}) }
  \\
=\sum_{\mu \in {\mathcal P}}
S_{[\mu]}(X|Y )S_{\mu }(T)
\sum_{\kappa \in {\mathcal P}^{-}} S_{\kappa}(T) 
\qquad [\text{by \eqref{co-B},  \eqref{sum-L2}}]
\\
=\sum_{\lambda , \mu \in {\mathcal P}}
\sum_{\kappa \in {\mathcal P}^{-}} 
 \mathrm{LR}^{\lambda}_{\kappa , \mu} 
S_{[\mu]}(X|Y )
S_{\lambda}(T) 
\quad [\text{by \eqref{LR}}] .
\label{dc-5ap}
\end{multline}
By comparing the coefficients of $S_{\lambda}(T)$ on both sides of \eqref{dc-5ap}, we obtain \eqref{dc-4}. 
Similarly, all other relations can be proven by modifying \eqref{co-A} using  
\eqref{LR}: 
\begin{itemize}
    \item Case \eqref{dc-1a}: using \eqref{co-B} and \eqref{sum-L1};
    \item Case \eqref{dc-1b}: using \eqref{co-C} and \eqref{sum-L2};
    \item Case \eqref{dc-2a}: using \eqref{co-B} and \eqref{sum-L2};
    \item Case \eqref{dc-2b}: using \eqref{co-B}, \eqref{sum-S} and \eqref{signS};
    \item Case \eqref{dc-3a}: using \eqref{co-C} and \eqref{sum-L1};
    \item Case \eqref{dc-3b}: using \eqref{co-C}, \eqref{sum-S} and \eqref{signS};
 \item Case \eqref{dc-4}: using \eqref{co-B} and \eqref{sum-L2};   
    \item Case \eqref{dc-4du}: using \eqref{co-C} and \eqref{sum-L1}.
\end{itemize}
Alternatively, one can derive \eqref{dc-1b} from \eqref{dc-1a},
\eqref{dc-3a} from \eqref{dc-2a}, 
\eqref{dc-3b} from \eqref{dc-2b}, and \eqref{dc-4du} from \eqref{dc-4} by using \eqref{LRtran}, 
\eqref{A-dual} and \eqref{BCD-dual}.
\end{proof}
%in case \eqref{dc-5b}, using \eqref{co-C} and \eqref{sum-L2}; 
%and in case \eqref{dc-5c}, using \eqref{co-B}, 
%\eqref{sum-S} and \eqref{signS}.
%%%%%%%%%%%%%%%%%%%%%%%%%%%%%%%%%

\section{Specialization}
In this section, we will consider specializations of \eqref{dc-1a}-\eqref{dc-4du}, and compare them with character 
formulas for Kirillov-Reshetikhin modules of quantum affine algebras (or their Yangian counterparts). 
We assume that the characters of a finite-type quantum group 
$U_{q}(\mathfrak{g})$ for generic $q$ coincide with those of the corresponding 
Lie (super)algebra $\mathfrak{g}$, and we use the same notation for modules of $U_{q}(\mathfrak{g})$ and those of $\mathfrak{g}$. 
For the non-super case, see for example 
Proposition~12 in \cite{KS97}. 
For the super case, we are not aware of a reference 
where this statement is explicitly formulated.
In the case of $\mathfrak{gl}(m|n)$, the same conclusion
follows from the crystal base theory of
$U_q(\mathfrak{gl}(m|n))$ developed by Benkart, Kang and Kashiwara
\cite{BKK98}.

We specialize the sets $X$ and $Y$ to  finite sets: 
$X=\{z_{1},z_{2},\dots , z_{M}\}$, 
$Y=\{z_{M+1},z_{M+2},\dots , z_{M+N}\}$. 
The generating function \eqref{genh} 
reduces to  
\begin{align}
\sum_{m=0}^{\infty} h_{m} (\{z_{b} \}_{b=1}^{M}  | \{z_{b+M}\}_{b=1}^{N} ) t^{m}=
\prod_{j=1}^{M} (1-z_{j} t)^{-1} \prod_{j=1}^{N} (1-z_{j+M} t) ,
\label{gench-A}
\end{align}
where 
$ h_{0} (\{z_{b} \}_{b=1}^{M}  | \{z_{b+M}\}_{b=1}^{N} )=1$, and $ h_{m} (\{z_{b} \}_{b=1}^{M}  | \{z_{b+M}\}_{b=1}^{N} )=0$ if $m<0$. 
 The determinant \eqref{JT-A} within the $[M,N]$-hook 
 (characterized by $\lambda_{M+N} \le N$) 
yields the supercharacter of the irreducible module 
$V(\Lambda)$ with the highest weight \eqref{HW-A}, as specified by \eqref{YW-A} \cite{BB81,BR87,VHKT90}.
 It also becomes zero if $\lambda$ is
  outside of the $[M,N]$-hook, namely $S_{\lambda}(\{z_{b} \}_{b=1}^{M}  | \{z_{b+M}\}_{b=1}^{N} )=0$ if $\lambda_{M+1}> N$. 
 We will consider several reductions (or foldings) of the 
supercharacters of $\mathfrak{gl}(M|N)$ associated with the Dynkin diagram symmetries 
of $\mathfrak{gl}(M|N)$.
Such symmetries arise in symmetric Dynkin diagrams. Nevertheless, we start from the supercharacters defined for the distinguished Dynkin diagram, because the supercharacters, as supersymmetric functions of $X$ and $Y$, are independent
\footnote{Eq.\ \eqref{YW-A} (thus the arrangement of rectangles in the Young diagram in Figure~\ref{MN-hookA}) depends on the choice of the Dynkin diagram; see~\cite{CW12}.
}
 of the choice of the Dynkin diagram.

We will consider the following specializations of the determinants \eqref{JT-B} and \eqref{JT-C} 
%(cf.\ \cite{BB81,BSR96})
:
\begin{align}
 &S_{[\lambda]} (\mathbf{x} \sqcup \{1\} \sqcup \mathbf{x}^{-1}|\mathbf{y} \sqcup \mathbf{y}^{-1})
& & \text{[type $\Bs$]},
\label{ch-fin-B}
\\
&  S_{[\lambda]} (\mathbf{x} \sqcup \mathbf{x}^{-1}|\mathbf{y} \sqcup \mathbf{y}^{-1})
& & \text{[type $\Ds$]},
\label{ch-fin-D}
\\
& S_{\langle \lambda \rangle }  (\mathbf{x}\sqcup \mathbf{x}^{-1}|\mathbf{y} \sqcup \mathbf{y}^{-1}) 
& &  \text{[type $\Cs$]} ,
\label{ch-fin-C}
\end{align}
where $\mathbf{x}=\{x_{b}\}_{b=1}^{r}$, $\mathbf{x}^{-1}=\{x_{b}^{-1}\}_{b=1}^{r}$, 
$\mathbf{y}=\{y_{b}\}_{b=1}^{s}$, $\mathbf{y}^{-1}=\{y_{b}^{-1}\}_{b=1}^{s}$. 
This type of determinant expressions for the (super)characters of ortho-symplectic Lie superalgebras appeared in \cite{BB81,BSR96}.
In \cite{CK87}, the corresponding (super)character formulas are defined through the Cauchy-type identity (3.31).
In addition, the relation between a Young diagram 
$\lambda $ (or partition) and the label of the representation is described in \cite{FJ84,MSS85}. 
In the case where the Young diagram $\lambda$ lies in the $[r,s]$-hook, \eqref{ch-fin-B} is expected to provide the supercharacters of 
$\mathfrak{osp}(2r+1|2s)$-modules whose highest weights are specified by 
(\eqref{HW-B}, \eqref{YW-B}) or (\eqref{HW-B0}, \eqref{YW-B0}). 
Similarly, \eqref{ch-fin-D} is expected to give the supercharacters of 
$\mathfrak{osp}(2r|2s)$-modules with highest weights specified by 
(\eqref{HW-D}, \eqref{YW-Dp}, \eqref{YW-Dm}) or 
(\eqref{HW-Cp}, \eqref{YW-Cp}).
 \footnote{
We do not know the general conditions under which these coincide with the {\em irreducible} supercharacter of $V_{\lambda}$. 
It may be necessary to subtract the supercharacters of unwanted 
invariant subspaces in order to obtain the irreducible ones.
}
Eq.~\eqref{ch-fin-C} is essentially equivalent to \eqref{ch-fin-D} in the sense that it can be expressed in terms of \eqref{ch-fin-D} through the relation \eqref{BCD-dual}:
\begin{align}
 S_{\langle \lambda \rangle }(\mathbf{x}\sqcup \mathbf{x}^{-1}\mid \mathbf{y}\sqcup \mathbf{y}^{-1})
   = (-1)^{|\lambda|}
      S_{[\lambda^{\prime}]}(\mathbf{y}\sqcup \mathbf{y}^{-1}\mid \mathbf{x}\sqcup \mathbf{x}^{-1}) .
\end{align}
Eq.~\eqref{ch-fin-C} is therefore expected to provide the supercharacters of 
$\mathfrak{spo}(2r|2s)$-modules, where the Lie superalgebra 
$\mathfrak{spo}(2r|2s)$ itself is isomorphic to $\mathfrak{osp}(2s|2r)$ 
under parity reversal.

In particular at $s=0$, 
$S_{[\lambda]} (\mathbf{x} \sqcup \{1\} \sqcup \mathbf{x}^{-1}|\emptyset)$, 
$S_{[\lambda]} (\mathbf{x} \sqcup \mathbf{x}^{-1}|\emptyset)$ and   
$S_{\langle \lambda \rangle }  (\mathbf{x} \sqcup \mathbf{x}^{-1}| \emptyset)$  
 give  irreducible characters of $\Bs_{r}=\mathfrak{so}(2r+1)$, $O(2r)$ and $\Cs_{r}=\mathfrak{sp}(2r)$, respectively. 
 In the case $\lambda_{r}=0$, 
$S_{[\lambda]}(\mathbf{x} \sqcup \mathbf{x}^{-1}\mid \emptyset)$ 
gives the irreducible character of the $\Ds_{r}=\mathfrak{so}(2r)$-module 
$V_{\lambda}$, whereas in the case $\lambda_{r}\neq 0$, 
it becomes the sum of the characters of the irreducible 
$\mathfrak{so}(2r)$-modules $V^{+}_{\lambda}$ and $V^{-}_{\lambda}$.
In this sense, \eqref{ch-fin-B}, \eqref{ch-fin-D}, and \eqref{ch-fin-C}
are generalizations of the character formulas for Lie algebras of types $\Bs$, $\Ds$, and $\Cs$, respectively.

Let us introduce the set of all the partitions 
$ \lambda =(\lambda_{1},\lambda_{2}, \dots , \lambda_{a})$ 
in a rectangular Young diagram $(m^{a})$ in the $[M,N]$-hook ($a,m \in \mathbb{Z}_{\ge 1}$; $m \le N$ if $a \ge M+1$):
\begin{align}
\ytableausetup{boxsize=0.5em}
\mathcal{S}_{\tiny \ydiagram{1}}=
\bigl\{ \lambda \ \bigl| \ 
&m \ge \lambda_{1} \ge \lambda_{2}  \ge \dots\ge \lambda_{a} \ge 0
\bigr\} ,
\label{req-dec}
\end{align}
and two kinds of subsets of this set:
\begin{align}
\mathcal{S}_{\tiny \ydiagram{1,1}}=
\Biggl\{ \lambda \ \Biggl| \ 
\begin{split}
&m = \lambda_{1} \ge \lambda_{2} =\lambda_{3} \ge \lambda_{4}=\lambda_{5} 
\ge \dots\ge \lambda_{a-1}=\lambda_{a} \ge 0 \quad \text{if} \quad a \in 2\mathbb{Z}+1,
\\
&m \ge \lambda_{1} = \lambda_{2} \ge \lambda_{3}=\lambda_{4} 
\ge \dots\ge \lambda_{a-1}=\lambda_{a} \ge 0 
\quad \text{if} \quad a \in 2\mathbb{Z}
\end{split}
\Biggr\} ,
\label{req-dect}
\\
\mathcal{S}_{\tiny \ydiagram{2}}=
\Biggl\{ \lambda \ \Biggl| \ 
\begin{split}
&m \ge \lambda_{1} \ge \lambda_{2}  \ge \dots\ge \lambda_{a} \ge 1, \quad \lambda_{j} \in 
2\mathbb{Z}+1
 \quad \text{if} \quad m \in 2\mathbb{Z}+1,
\\
&m \ge \lambda_{1} \ge \lambda_{2} \ge \dots\ge \lambda_{a} \ge 0 , \quad \lambda_{j} \in 
2\mathbb{Z} 
\quad \text{if} \quad m \in 2\mathbb{Z}
\end{split}
\Biggr\} .
\label{req-decy}
\end{align}
%%%%%%%%%%%%%%%%%%%%
It is known \cite{KR90,Ki89,HKOTY98,HKOTT01} that the characters of Kirillov-Reshetikhin modules over quantum affine algebras can be expressed as linear combinations of the characters of their finite-type subalgebras, reflecting the decomposition \eqref{KRdecom}.
The subset we consider in this paper has, under our convention, the following form for $m \in \mathbb{Z}_{\ge 1}$.

\noindent For $U_{q}(\mathfrak{so}(2r+1)^{(1)})$, $a \in \{1,2,\dots , r\}$:
\begin{align}
\ytableausetup{boxsize=0.5em}
\mathrm{ch} W^{(a)}_{(1+\delta_{a r})m}
&=\sum_{\lambda \in \mathcal{S}_{\tiny \ydiagram{1,1}}} S_{[\lambda]} (\mathbf{x} \sqcup \{1\} \sqcup \mathbf{x}^{-1}|\emptyset ) 
&& \text{[on type $\Bs$]}
\label{de-BB0}
\\
&=\sum_{\lambda \in \mathcal{S}_{\tiny \ydiagram{1}}} S_{[\lambda]} (\mathbf{x} \sqcup \mathbf{x}^{-1}|\emptyset ) 
&& \text{[on type $\Ds$]}.
\label{de-BD0}
\end{align}
For $U_{q}(\mathfrak{sl}(2r+1)^{(2)})$, $a \in \{1,2,\dots , r\}$:
\begin{align}
\ytableausetup{boxsize=0.5em}
\mathrm{ch} W^{(a)}_{m}
&=\sum_{\lambda \in \mathcal{S}_{\tiny \ydiagram{2}}} S_{[\lambda]} (\mathbf{x} \sqcup \{1\} \sqcup \mathbf{x}^{-1}|\emptyset ) 
&& \text{[on type $\Bs$]}
\label{de-t2B0}
\\
&=\sum_{\lambda \in \mathcal{S}_{\tiny \ydiagram{1}}} S_{\langle \lambda \rangle} (\mathbf{x} \sqcup \mathbf{x}^{-1}|\emptyset ) 
&& \text{[on type $\Cs$]}.
\label{de-t2C0}
\end{align}
For $U_{q}(\mathfrak{sl}(2r)^{(2)})$, $a \in \{1,2,\dots , r\}$:
\begin{align}
\ytableausetup{boxsize=0.5em}
\mathrm{ch} W^{(a)}_{m}
&=\sum_{\lambda \in \mathcal{S}_{\tiny \ydiagram{2}}} S_{[\lambda]} (\mathbf{x} \sqcup \mathbf{x}^{-1}|\emptyset ) 
&& \text{[on type $\Ds$]}
\label{de-t3D0}
\\
&=\sum_{\lambda \in \mathcal{S}_{\tiny \ydiagram{1,1}}} S_{\langle \lambda \rangle} (\mathbf{x} \sqcup \mathbf{x}^{-1}|\emptyset) 
&& \text{[on type $\Cs$]}.
\label{de-t3C0}
\end{align}
For $U_{q}(\mathfrak{so}(2r)^{(1)})$, $a \in \{1,2,\dots , r-2\}$:
\begin{align}
\ytableausetup{boxsize=0.5em}
\mathrm{ch} W^{(a)}_{m}
&=\sum_{\lambda \in \mathcal{S}_{\tiny \ydiagram{1,1}}} S_{[\lambda]} (\mathbf{x} \sqcup \mathbf{x}^{-1}|\emptyset) 
&& \text{[on type $\Ds$]} .
\label{de-D0}
\end{align}
For $U_{q}(\mathfrak{sp}(2r)^{(1)})$, $a \in \{1,2,\dots , r-1\}$:
\begin{align}
\ytableausetup{boxsize=0.5em}
\mathrm{ch} W^{(a)}_{m} &=
\sum_{\lambda \in \mathcal{S}_{\tiny \ydiagram{2}}}
S_{\langle \lambda \rangle}(\mathbf{x} \sqcup \mathbf{x}^{-1} |\emptyset) 
&& \text{[on type $\Cs$]} .
\label{de-C0}
\end{align}
For $U_{q}(\mathfrak{so}(2r+2)^{(2)})$, $a \in \{1,2,\dots , r-1\}$:
\begin{align}
\ytableausetup{boxsize=0.5em}
\mathrm{ch} W^{(a)}_{m}
&=\sum_{\lambda \in \mathcal{S}_{\tiny \ydiagram{1}}} S_{[\lambda]} (\mathbf{x}\sqcup \{1\}\sqcup \mathbf{x}^{-1}|\emptyset) 
&& \text{[on type $\Bs$]}.
\label{de-D20}
\end{align}
Here we do not deal with the spinor-type representations 
$W^{(r)}_{2m-1}$ for $U_{q}(\mathfrak{so}(2r+1)^{(1)})$, 
$W^{(r-1)}_{m}$ and $W^{(r)}_{m}$ for $U_{q}(\mathfrak{so}(2r)^{(1)})$, 
$W^{(r)}_{m}$ for $U_{q}(\mathfrak{so}(2r+2)^{(2)})$, and 
the representation 
$W^{(r)}_{m}$ for $U_{q}(\mathfrak{sp}(2r)^{(1)})$
\footnote{Note, however, that the character of
$W^{(r)}_{2m-1}$ for $U_{q}(\mathfrak{so}(2r+1)^{(1)})$ 
is described as an asymptotic limit of the supercharacter of a typical representation of $\mathfrak{gl}(2r|1)$ \cite{T21} (see also \cite{T11}); the characters of 
$W^{(r-1)}_{m}$ and $W^{(r)}_{m}$ for $U_{q}(\mathfrak{so}(2r)^{(1)})$ 
 coincide with those of $\mathfrak{so}(2r)$ \cite{KR90,HKOTY98}; 
 the character of  
$W^{(r)}_{m}$ for $U_{q}(\mathfrak{so}(2r+2)^{(2)})$ coincides with 
that of $\mathfrak{so}(2r+1)$ \cite{HKOTT01}; the 
character of $W^{(r)}_{m}$ for $U_{q}(\mathfrak{sp}(2r)^{(1)})$ 
coincides with that of $\mathfrak{sp}(2r)$ \cite{KR90,HKOTY98}. 
In total, the characters of the Kirillov-Reshetikhin modules discussed here can be obtained solely from those of finite-dimensional Lie (super)algebras or their suitable specializations, without taking linear combinations of characters in the conventional sense.}
. 
Our claim is that all of these characters can be obtained
as folding (or reduction) of supercharacters of the 
finite-dimensional Lie superalgebra $\mathfrak{gl}(M|N)$.
%%%%%%%%%%%%%%
\begin{theorem}
\label{main-th}
The character formulas \eqref{de-BB0}--\eqref{de-D20} 
can be expressed in terms of supersymmetric Schur functions as follows.
\begin{flalign}
& U_{q}(\mathfrak{so}(2r+1)^{(1)}), \quad a \in \{1,2,\dots , r\}: 
& 
& \mathrm{ch}\, W^{(a)}_{(1+\delta_{ar})m}
=
S_{(m^{a})}(\mathbf{x} \sqcup \mathbf{x}^{-1} \mid \{-1\}), &
\label{de-BB00}
\end{flalign}

\begin{flalign}
& U_{q}(\mathfrak{sl}(2r+1)^{(2)}), \quad a \in \{1,2,\dots , r\}:
&
& \mathrm{ch}\, W^{(a)}_{m}
=
S_{(m^{a})}(\mathbf{x} \sqcup \{1\} \sqcup \mathbf{x}^{-1} \mid \emptyset), &
\label{de-t2B00}
\end{flalign}

\begin{flalign}
& U_{q}(\mathfrak{sl}(2r)^{(2)}), \quad a \in \{1,2,\dots , r\}:
&
& \mathrm{ch}\, W^{(a)}_{m}
=
S_{(m^{a})}(\mathbf{x} \sqcup \mathbf{x}^{-1} \mid \emptyset), &
\label{de-t3D00}
\end{flalign}

\begin{flalign}
& U_{q}(\mathfrak{so}(2r)^{(1)}), \quad a \in \{1,2,\dots , r-2\}:
&
& \mathrm{ch}\, W^{(a)}_{m}
=
S_{(m^{a})}(\mathbf{x} \sqcup \mathbf{x}^{-1} \mid \{1,-1\}), &
\label{de-D00}
\end{flalign}

\begin{flalign}
& U_{q}(\mathfrak{sp}(2r)^{(1)}), \quad a \in \{1,2,\dots , r-1\}:
&
& \mathrm{ch}\, W^{(a)}_{m}
=
S_{(m^{a})}(\mathbf{x} \sqcup \{1,-1\} \sqcup \mathbf{x}^{-1} \mid \emptyset), &
\label{de-C00}
\end{flalign}

\begin{flalign}
& U_{q}(\mathfrak{so}(2r+2)^{(2)}), \quad a \in \{1,2,\dots , r-1\}:
&
& \mathrm{ch}\, W^{(a)}_{m}
=
S_{(m^{a})}(\mathbf{x} \sqcup \{1,1\} \sqcup \mathbf{x}^{-1} \mid \{1,-1\}) &
\nonumber\\
& &
& =
S_{(m^{a})}(\mathbf{x} \sqcup \{1\} \sqcup \mathbf{x}^{-1} \mid \{-1\}). &
\label{de-D200}
\end{flalign}
\end{theorem}
%%%%%%%%%%%%%%%%%%
\begin{proof}
It suffices to show that, in each case, the right-hand side of \eqref{de-BB00}--\eqref{de-D200} coincides with the corresponding equation in \eqref{de-BB0}--\eqref{de-D20}. This follows from the appropriate specialization of Proposition~\ref{pro-gen}, combined with the identities: 
\begin{align}
\ytableausetup{boxsize=0.5em}
\sum_{\nu \in {\mathcal P}}
 \mathrm{LR}^{(m^{a})}_{\nu , \mu} 
&=\begin{cases}
1 & \text{if} \quad \mu \in \mathcal{S}_{\tiny \ydiagram{1}} \\
0 & \text{otherwise},
\end{cases}
\label{LRrq1} 
\\
\sum_{\kappa \in {\mathcal P}^{-} } \mathrm{LR}^{(m^{a})}_{\kappa , \mu} 
&=
\begin{cases}
1 & \text{if} \quad \mu \in \mathcal{S}_{\tiny \ydiagram{1,1}} \\
0 & \text{otherwise},
\end{cases}
\label{LRrq2}
\\
\sum_{\kappa \in {\mathcal P}^{+} } \mathrm{LR}^{(m^{a})}_{\kappa , \mu} 
&=
\begin{cases}
1 & \text{if} \quad \mu \in \mathcal{S}_{\tiny \ydiagram{2}} \\
0 & \text{otherwise},
\end{cases}
\label{LRrq3}
\end{align}
which are consequences of \eqref{LR5}. 

Specifically:

\medskip
\noindent
{\bf (i) $U_q(\mathfrak{so}(2r+1)^{(1)})$.}
By setting 
\[
X=\mathbf{x} \sqcup \mathbf{x}^{-1}, 
\quad Y=\emptyset , \quad \xi=-1, \quad \lambda=(m^{a}) 
\]
in \eqref{dc-2a} and applying \eqref{LRrq2}, one can verify that the right-hand side of \eqref{de-BB00} coincides with that of \eqref{de-BB0}.
%%%%

\medskip
\noindent
{\bf (ii) $U_q(\mathfrak{sl}(2r+1)^{(2)})$.}
By setting 
\[
X=\mathbf{x} \sqcup \{1 \} \sqcup \mathbf{x}^{-1}, 
\quad Y=\emptyset , \quad \lambda=(m^{a}) 
\]
in \eqref{dc-1a} and applying \eqref{LRrq3}, one can verify that the right-hand side of \eqref{de-t2B00} coincides with that of \eqref{de-t2B0}.
%%%%

\medskip
\noindent
{\bf (iii) $U_q(\mathfrak{sl}(2r)^{(2)})$.}
By setting 
\[
X=\mathbf{x} \sqcup \mathbf{x}^{-1}, 
\quad Y=\emptyset , \quad \lambda=(m^{a}) 
\]
in \eqref{dc-1a} and applying \eqref{LRrq3}, 
one can verify that the right-hand side of \eqref{de-t3D00} coincides with that of \eqref{de-t3D0}.

\medskip
\noindent
{\bf (iv) $U_q(\mathfrak{so}(2r)^{(1)})$.}
By setting 
\[
X=\mathbf{x} \sqcup \mathbf{x}^{-1}, 
\quad Y=\emptyset , \quad \lambda=(m^{a}) 
\]
in \eqref{dc-4} and applying \eqref{LRrq2}
 , one can verify that the right-hand side of \eqref{de-D00}
  coincides with that of \eqref{de-D0}.

%%%%
\medskip
\noindent
{\bf (v) $U_q(\mathfrak{sp}(2r)^{(1)})$.}
By setting 
\[
X=\mathbf{x} \sqcup \mathbf{x}^{-1}, 
\quad Y=\emptyset , \quad \lambda=(m^{a}) 
\]
in \eqref{dc-4du} and applying \eqref{LRrq3}, 
 one can verify that the right-hand side of \eqref{de-C00}
  coincides with that of \eqref{de-C0}.

%%%%
\medskip
\noindent
{\bf (vi) $U_q(\mathfrak{so}(2r+2)^{(2)})$.}
By setting 
\[
X=\mathbf{x} \sqcup \{1,1 \} \sqcup \mathbf{x}^{-1}, 
\quad Y=\{1\} , \quad \xi=-1, \quad \lambda=(m^{a}) 
\]
in \eqref{dc-2b} and applying \eqref{SP-A} and \eqref{LRrq1}
 , one can verify that the right-hand side of \eqref{de-D200}
  coincides with that of \eqref{de-D20}.
\end{proof}
%%%%%%%%%%%%%
We remark that the folding technique for q-characters can be found in \cite{KOSY01} in the context of \eqref{de-C00}, while the folding technique for transfer-matrix eigenvalues can be found in \cite{KS94-2} in the context of \eqref{de-t2B00} and \eqref{de-t3D00}, in \cite{T21,T11} in the context of \eqref{de-BB00}
\footnote{A representation-theoretical background relevant to this is provided in \cite{BM14}.}, and in \cite{T23} for the superalgebra cases that include all of these \eqref{de-BB00}-\eqref{de-D200}. 
Note that, once the dependence on the spectral parameter is appropriately dropped, the q-characters or transfer-matrix eigenvalues reduce to ordinary characters.
In the following, we present the corresponding formulas in a more general setting, namely for representations of quantum affine superalgebras labeled by arbitrary Young diagrams, where the module 
$V(\Lambda)$ in \eqref{KRdecom} corresponds to the module $V_{\lambda}$ defined in Section~2.
This provides a proof of the conjectures stated in Appendix~B of \cite{T23}.

%%%%%%%%%%%%%%%%%%%%%%%%%%%%%%%%%%%%%%%%
\paragraph{$U_{q}(\mathfrak{osp}(2r+1|2s)^{(1)})$ case:}
Specializing the sets $X$ and $Y$ in \eqref{genh} 
to  finite sets: 
$X=\mathbf{x}\sqcup \mathbf{x}^{-1}$ and 
$Y=\mathbf{y}\sqcup \{-1\} \sqcup \mathbf{y}^{-1}$,  
we obtain 
\begin{multline}
%w(t)&=
\prod_{j=1}^{r} (1-x_{j} t)^{-1} (1-x_{j}^{-1} t)^{-1} \prod_{j=1}^{s} (1-y_{j} t)(1-y_{j}^{-1} t)(1+t)=
% \nonumber \\
%
%&
\\
=\sum_{m=0}^{\infty} h_{m}(\mathbf{x} \sqcup \mathbf{x}^{-1}|\mathbf{y}\sqcup \{-1\} \sqcup \mathbf{y}^{-1}) t^{m} .
\label{gench-B}
\end{multline}
This corresponds to the folding of the $\mathfrak{gl}(2r|2s+1)$  case in  \eqref{gench-A}, under the identifications 
$x_{j}=z_{j}=z_{2r+1-j}^{-1}$ for $1 \le j \le r$, 
$y_{j}=z_{2r+j}=z_{2r+2s+2-j}^{-1}$ for $1 \le j \le s$ and  
$z_{2r+s+1}=-1$. 

Applying 
\eqref{dc-2a} and \eqref{dc-2b} for $\xi =-1$ to $S_{\lambda}(X|Y)$, we obtain 
\begin{align}
S_{\lambda}(\mathbf{x}\sqcup \mathbf{x}^{-1}|\mathbf{y}\sqcup \{-1\} \sqcup \mathbf{y}^{-1})&=\sum_{\mu \in {\mathcal P}}
\sum_{\kappa \in {\mathcal P}^{-} } \mathrm{LR}^{\lambda}_{\kappa , \mu} 
S_{[\mu]}(\mathbf{x}\sqcup \{1\} \sqcup \mathbf{x}^{-1}|\mathbf{y} \sqcup \mathbf{y}^{-1})
\label{dc-2aB}
\\
&=\sum_{\mu ,\nu \in {\mathcal P}}
 \mathrm{LR}^{\lambda}_{\nu , \mu} 
S_{[ \mu ]}(\mathbf{x}\sqcup \mathbf{x}^{-1} |\mathbf{y} \sqcup \mathbf{y}^{-1}).
\label{dc-2bB}
\end{align}
Note that this is non-trivial only if $\lambda$ is in the 
$[2r,2s+1]$-hook. 
A proof of \eqref{dc-2aB} restricted to the $[r,0]$-hook for the case $s=0$ is available in  [Lemma 7.3, \cite{HKOTY98}] (see also [eq.\ (3.14), \cite{KOS95}]).  

Specializing \eqref{dc-2aB} and \eqref{dc-2bB} to 
 a rectangular Young diagram $\lambda=(m^{a})$ in the $[2r,2s+1]$-hook ($a,m \in \mathbb{Z}_{\ge 1}$), 
 and using \eqref{LRrq2} and \eqref{LRrq1}, we obtain the following
 decomposition formulas:
\begin{align}
\ytableausetup{boxsize=0.5em}
S_{(m^{a})}(\mathbf{x} \sqcup \mathbf{x}^{-1}|\mathbf{y}\sqcup \{-1\} \sqcup \mathbf{y}^{-1})
&=\sum_{\lambda \in \mathcal{S}_{\tiny \ydiagram{1,1}}} S_{[\lambda]} (\mathbf{x} \sqcup \{1\} \sqcup \mathbf{x}^{-1}|\mathbf{y}\sqcup \mathbf{y}^{-1}) 
&& \text{[on type $\Bs$]}
\label{de-BB}
\\
&=\sum_{\lambda \in \mathcal{S}_{\tiny \ydiagram{1}}} S_{[\lambda]} (\mathbf{x} \sqcup \mathbf{x}^{-1}|\mathbf{y}\sqcup \mathbf{y}^{-1}) 
&& \text{[on type $\Ds$]}.
\label{de-BD}
\end{align}
To relate these formulas to the labels of representations 
defined in Section 2, we  restrict them to the $[r,s]$-hook (thus $m \le s$ if $a\ge r+1$).  
In this case, our observation in \cite{T23} on the 
Bethe strap \cite{KS94-1,Su95} suggests that they yield irreducible supercharacters of $U_{q}(\mathfrak{osp}(2r+1|2s)^{(1)})$. 
 Eq.\ \eqref{de-BB} suggests a decomposition
\footnote{Compare this with eq.~(4.30) in \cite{T99} for the case $r=0$.}
  of a
 module $W_{a,m}$ of $U_{q}(\mathfrak{osp}(2r+1|2s)^{(1)})$ (or $Y(\mathfrak{osp}(2r+1|2s))$) 
into modules $\{V^{\prime}_{\lambda }\}$ of $U_{q}(\mathfrak{osp}(2r+1|2s))$ (or $\mathfrak{osp}(2r+1|2s)$): 
$W_{a,m} \simeq
\oplus_{\lambda \in \mathcal{S}_{\tiny \ydiagram{1,1}}} V^{\prime}_{\lambda}  $.  
%Here we use the same symbol $\lambda$ for a Young
% diagram and the highest weight specified by it 
% (via \eqref{HW-B}, \eqref{YW-B}). 
We do not know whether 
 $V^{\prime}_{\lambda}$ coincides with the irreducible
  module  $V_{\lambda}$ 
 in the general situation, but at least for $s=0$ it does. 
 In this case ($s=0$), \eqref{de-BB} coincides with the 
character formula of the Kirillov-Reshetikhin module  
$W_{a,m}=W^{(a)}_{(1+\delta_{a r})m}$ \cite{KR90} 
for $U_{q}(\mathfrak{so}(2r+1)^{(1)})$ or $Y(\mathfrak{so}(2r+1))$ (see also \cite{HKOTY98}; 
use \eqref{KD-B} for comparison; 
eq.\eqref{de-BB} yields \eqref{de-BB0} and \eqref{de-BB00}
 upon substituting $s = 0$). Note that the case $a=r$ (with $s=0$) 
corresponds to a spin-even 
(tensor-like) representation $W^{(r)}_{2m}$, and the spin-odd (spinor-like) representations  $W^{(r)}_{2m-1}$ have 
to be treated separately.
The second equality \eqref{de-BD} for $s=0$ corresponds to [eq.\ (C.1), \cite{HKOTT01}], 
which suggests another decomposition of 
$W^{(a)}_{(1+\delta_{a r})m}$ (eq.\ \eqref{de-BD} 
yields \eqref{de-BD0} and \eqref{de-BB00}
 upon substituting $s = 0$). 

%$S_{a,m}=S_{(m^{a})}(\mathbf{x} \sqcup \mathbf{x}^{-1}|\mathbf{y},-1,\mathbf{y}^{-1})$ 
%for $1 \le a \le r-1$, 
%$S_{r,2m}=S_{(m^{r})}(\mathbf{x} \sqcup \mathbf{x}^{-1}|\mathbf{y},-1,\mathbf{y}^{-1})$.  

%%%%%%%%%%%%%%%%%%%%%%%%%%%%%%%%%%%%%%%%
\paragraph{$U_{q}(\mathfrak{sl}(2r|2s+1)^{(2)})$ case:}
Specializing the sets $X$ and $Y$ in \eqref{genh} 
to  finite sets: 
$X=\mathbf{x}\sqcup \mathbf{x}^{-1}$, 
$Y=\mathbf{y}\sqcup \{1\} \sqcup \mathbf{y}^{-1}$, 
we obtain
\begin{multline}
%w(t)&=
\prod_{j=1}^{r} (1-x_{j} t)^{-1} (1-x_{j}^{-1} t)^{-1} \prod_{j=1}^{s} (1-y_{j} t)(1-y_{j}^{-1} t)(1-t)=
% \nonumber \\
%
%&
\\
=\sum_{m=0}^{\infty} h_{m}(\mathbf{x} \sqcup \mathbf{x}^{-1}|\mathbf{y}\sqcup \{1\} \sqcup \mathbf{y}^{-1}) t^{m} .
\label{gench-t1}
\end{multline}
This corresponds to the folding of the $\mathfrak{gl}(2r|2s+1)$  case in  \eqref{gench-A}, under the identifications 
$x_{j}=z_{j}=z_{2r+1-j}^{-1}$ for $1 \le j \le r$, 
$y_{j}=z_{2r+j}=z_{2r+2s+2-j}^{-1}$ for $1 \le j \le s$ and  
$z_{2r+s+1}=1$. 

Applying \eqref{dc-2a} and \eqref{dc-2b} for $\xi =1$ to $S_{\lambda}(X|Y)$, we obtain
\begin{align}
S_{\lambda}(\mathbf{x}\sqcup \mathbf{x}^{-1}|\mathbf{y}\sqcup \{1\} \sqcup \mathbf{y}^{-1})&=\sum_{\mu \in {\mathcal P}}
\sum_{\kappa \in {\mathcal P}^{-} } \mathrm{LR}^{\lambda}_{\kappa , \mu} 
S_{[\mu]}(\mathbf{x}\sqcup \{-1\} \sqcup \mathbf{x}^{-1} |\mathbf{y} \sqcup \mathbf{y}^{-1})
\label{dc-2at1}
\\
&=\sum_{\mu,\nu \in {\mathcal P}}
(-1)^{|\nu|} \mathrm{LR}^{\lambda}_{\nu , \mu} 
S_{[\mu]}(\mathbf{x}\sqcup \mathbf{x}^{-1} |\mathbf{y} \sqcup \mathbf{y}^{-1}).
\label{dc-2bt1}
\end{align}
Note that this is non-trivial only if $\lambda$ is in the 
$[2r,2s+1]$-hook. 
Specializing \eqref{dc-2at1} and \eqref{dc-2bt1} to a
 rectangular Young diagram $(m^{a})$ in the $[2r,2s+1]$-hook ($a,m \in \mathbb{Z}_{\ge 1}$), and using \eqref{LRrq2} and \eqref{LRrq1}, we obtain the following decomposition formulas:
\begin{align}
\ytableausetup{boxsize=0.5em}
S_{(m^{a})}(\mathbf{x} \sqcup \mathbf{x}^{-1}|\mathbf{y}\sqcup \{1\} \sqcup \mathbf{y}^{-1})
&=\sum_{\lambda \in \mathcal{S}_{\tiny \ydiagram{1,1}}} S_{[\lambda]} (\mathbf{x} \sqcup \{-1\} \sqcup \mathbf{x}^{-1}|\mathbf{y}\sqcup \mathbf{y}^{-1}) 
&&  \text{[on type $\Bs$$^{\prime}$] }
\label{de-t1Bp}
\\
&=\sum_{\lambda \in \mathcal{S}_{\tiny \ydiagram{1}}} (-1)^{ma+|\lambda |}
 S_{[\lambda]} (\mathbf{x} \sqcup \mathbf{x}^{-1}|\mathbf{y}\sqcup \mathbf{y}^{-1}) 
&& \text{[on type $\Ds$]},
\label{de-t1Dp}
\end{align}
where $|\lambda|=\sum_{j=1}^{a}\lambda_{j}$ is the size of the Young diagram $\lambda$. 
This case is parallel to the case $U_{q}(\mathfrak{osp}(2r+1|2s)^{(1)})$. However, 
because of the difference between the factors $1+t$ in
 \eqref{gench-B} and $1-t$ in \eqref{gench-t1}, 
we have to modify \eqref{ch-fin-B} as in \eqref{de-t1Bp}, or add a sign factor as in \eqref{de-t1Dp}. 

%%%%%%%%%%%%%%%%%%%%%%%%%%%%%%%%%%%%%%%%
\paragraph{$U_{q}(\mathfrak{sl}(2r+1|2s)^{(2)})$ case:}
Specializing the sets $X$ and $Y$ in \eqref{genh} 
to  finite sets: 
$X=\mathbf{x}\sqcup \{1\} \sqcup \mathbf{x}^{-1}$, 
$Y=\mathbf{y}\sqcup  \mathbf{y}^{-1}$,  
we obtain  
\begin{multline}
%w(t)&=
\prod_{j=1}^{r} (1-x_{j} t)^{-1} (1-x_{j}^{-1} t)^{-1} \prod_{j=1}^{s} (1-y_{j} t)(1-y_{j}^{-1} t)(1-t)^{-1} =
\\
=\sum_{m=0}^{\infty} h_{m}(\mathbf{x} \sqcup \{1\} \sqcup \mathbf{x}^{-1}|\mathbf{y}\sqcup \mathbf{y}^{-1}) t^{m} .
\label{gench-t2}
\end{multline}
This corresponds to the folding of the $\mathfrak{gl}(2r+1|2s)$  case in  \eqref{gench-A}, under the identifications 
$x_{j}=z_{j}=z_{2r+2-j}^{-1}$ for $1 \le j \le r$, 
$y_{j}=z_{2r+1+j}=z_{2r+2s+2-j}^{-1}$ for $1 \le j \le s$ and  
$z_{2r+1}=1$. 

Applying \eqref{dc-1a} and \eqref{dc-3b} for $\xi =1$ to $S_{\lambda}(X|Y)$, we obtain 
\begin{align}
S_{\lambda}(\mathbf{x}\sqcup \{1\} \sqcup \mathbf{x}^{-1} |\mathbf{y}\sqcup  \mathbf{y}^{-1})&=\sum_{\mu \in {\mathcal P}}
\sum_{\kappa \in {\mathcal P}^{+} } \mathrm{LR}^{\lambda}_{\kappa , \mu} 
S_{[\mu]}(\mathbf{x}\sqcup \{1\} \sqcup \mathbf{x}^{-1} |\mathbf{y}\sqcup  \mathbf{y}^{-1})
\label{dc-1at2}
\\
&=\sum_{\mu,\nu \in {\mathcal P}}
 \mathrm{LR}^{\lambda}_{\nu , \mu} 
S_{\langle \mu \rangle}(\mathbf{x} \sqcup \mathbf{x}^{-1} |\mathbf{y}\sqcup  \mathbf{y}^{-1}).
\label{dc-3bt2}
\end{align}
Note that this is non-trivial only if $\lambda$ is in the 
$[2r+1,2s]$-hook. 
Specializing \eqref{dc-1at2} and \eqref{dc-3bt2} to a
 rectangular Young diagram $(m^{a})$ in the $[2r+1,2s]$-hook ($a,m \in \mathbb{Z}_{\ge 1}$), and using \eqref{LRrq3} and \eqref{LRrq1}, we obtain the following decomposition formulas:
\begin{align}
\ytableausetup{boxsize=0.5em}
S_{(m^{a})}(\mathbf{x} \sqcup \{1\} \sqcup \mathbf{x}^{-1}|\mathbf{y} 
\sqcup \mathbf{y}^{-1})
&=\sum_{\lambda \in \mathcal{S}_{\tiny \ydiagram{2}}} S_{[\lambda]} (\mathbf{x} \sqcup \{1\} \sqcup \mathbf{x}^{-1}|\mathbf{y} \sqcup  \mathbf{y}^{-1}) 
&& \text{[on type $\Bs$]}
\label{de-t2B}
\\
&=\sum_{\lambda \in \mathcal{S}_{\tiny \ydiagram{1}}} S_{\langle \lambda \rangle} (\mathbf{x} \sqcup \mathbf{x}^{-1}|\mathbf{y}\sqcup \mathbf{y}^{-1}) 
&& \text{[on type $\Cs$]}.
\label{de-t2C}
\end{align}
To relate these formulas to the labels of representations, we  restrict them to the $[r,s]$-hook. 
%In this case, our observation on the Bethe strap suggests that  it gives an
% irreducible character of $U_{q}(\mathfrak{sl}(2r+1|2s)^{(2)})$. 
 Eq.\  \eqref{de-t2B} suggests a decomposition of a
  module $W_{a,m}$ of $U_{q}(\mathfrak{sl}(2r+1|2s)^{(2)})$ 
into modules $\{V^{\prime}_{\lambda}\}$ of $U_{q}(\mathfrak{osp}(2r+1|2s))$: $W_{a,m} \simeq
\oplus_{\lambda \in \mathcal{S}_{\tiny \ydiagram{2}}} V^{\prime}_{\lambda}  $.  
In particular, 
 $V^{\prime}_{\lambda}$ coincides with the irreducible module   $V_{\lambda}$ 
 at least for $s=0$, and then \eqref{de-t2B} coincides with the 
character formula of the Kirillov-Reshetikhin module 
$W_{a,m}=W^{(a)}_{m}$ for 
$U_{q}(\mathfrak{sl}(2r+1)^{(2)})$ [eq.\ (6.7), \cite{HKOTT01}] 
(eq.\ \eqref{de-t2B} yields \eqref{de-t2B0} and \eqref{de-t2B00}
  upon substituting $s = 0$).  
The second equality \eqref{de-t2C} for $s=0$ corresponds to [eq.\ (6.6), \cite{HKOTT01}]
(eq.\ \eqref{de-t2C} yields \eqref{de-t2C0} and \eqref{de-t2B00}
  upon substituting $s = 0$), 
which suggests another decomposition of $W^{(a)}_{m}$. 

%%%%%%%%%%%%%%%%%%%%%%%%%%%%%%%%%%%%%%%%
\paragraph{$U_{q}(\mathfrak{sl}(2r|2s)^{(2)})$ case:}
Specializing the sets $X$ and $Y$ in \eqref{genh} 
to  finite sets: 
$X=\mathbf{x} \sqcup \mathbf{x}^{-1}$, 
$Y=\mathbf{y}\sqcup  \mathbf{y}^{-1}$,  
we obtain  
\begin{align}
%w(t)&=
\prod_{j=1}^{r} (1-x_{j} t)^{-1} (1-x_{j}^{-1} t)^{-1} \prod_{j=1}^{s} (1-y_{j} t)(1-y_{j}^{-1} t)
% \nonumber \\
%
%&
=\sum_{m=0}^{\infty} h_{m}(\mathbf{x} \sqcup \mathbf{x}^{-1}|\mathbf{y}\sqcup \mathbf{y}^{-1}) t^{m} .
\label{gench-t3}
\end{align}
This corresponds to the folding of the $\mathfrak{gl}(2r|2s)$  case in  \eqref{gench-A}, under the identifications 
$x_{j}=z_{j}=z_{2r+1-j}^{-1}$ for $1 \le j \le r$ and  
$y_{j}=z_{2r+j}=z_{2r+2s+1-j}^{-1}$ for $1 \le j \le s$. 
Applying \eqref{dc-1a} and \eqref{dc-1b} to $S_{\lambda}(X|Y)$, we obtain 
\begin{align}
S_{\lambda}(\mathbf{x} \sqcup \mathbf{x}^{-1}|\mathbf{y}\sqcup  \mathbf{y}^{-1})&=\sum_{\mu \in {\mathcal P}}
\sum_{\kappa \in {\mathcal P}^{+} } \mathrm{LR}^{\lambda}_{\kappa , \mu} 
S_{[\mu]}(\mathbf{x} \sqcup \mathbf{x}^{-1}|\mathbf{y}\sqcup  \mathbf{y}^{-1})
\label{dc-1at3}
\\
&=\sum_{\mu \in {\mathcal P}}
\sum_{\kappa \in {\mathcal P}^{-} }
 \mathrm{LR}^{\lambda}_{\kappa , \mu} 
S_{\langle \mu \rangle}(\mathbf{x} \sqcup \mathbf{x}^{-1} |\mathbf{y}\sqcup  \mathbf{y}^{-1}).
\label{dc-1bt3}
\end{align}
Note that this is non-trivial only if $\lambda$ is in the 
$[2r,2s]$-hook. 
Specializing \eqref{dc-1at3} and \eqref{dc-1bt3} to a
 rectangular Young diagram $(m^{a})$ in the $[2r,2s]$-hook ($a,m \in \mathbb{Z}_{\ge 1}$), and using \eqref{LRrq3} and \eqref{LRrq2}, we obtain the following decomposition formulas:
\begin{align}
\ytableausetup{boxsize=0.5em}
S_{(m^{a})}(\mathbf{x} \sqcup \mathbf{x}^{-1}|\mathbf{y}\sqcup \mathbf{y}^{-1})
&=\sum_{\lambda \in \mathcal{S}_{\tiny \ydiagram{2}}} S_{[\lambda]} (\mathbf{x} \sqcup \mathbf{x}^{-1}|\mathbf{y}\sqcup \mathbf{y}^{-1}) 
&& \text{[on type $\Ds$]}
\label{de-t3D}
\\
&=\sum_{\lambda \in \mathcal{S}_{\tiny \ydiagram{1,1}}} S_{\langle \lambda \rangle} (\mathbf{x} \sqcup \mathbf{x}^{-1}|\mathbf{y} \sqcup \mathbf{y}^{-1}) 
&& \text{[on type $\Cs$]}.
\label{de-t3C}
\end{align}
To relate these formulas to the labels of representations, we  restrict them to the $[r,s]$-hook. 
%In this case, our observation on the Bethe strap suggests that  it gives an
% irreducible character of $U_{q}(\mathfrak{sl}(2r|2s)^{(2)})$.
Eq.\  \eqref{de-t3D} suggests a decomposition of a module $W_{a,m}$ of $U_{q}(\mathfrak{sl}(2r|2s)^{(2)})$ 
into modules $\{V^{\prime}_{\lambda}\}$ of $U_{q}(\mathfrak{osp}(2r|2s))$: $W_{a,m} \simeq
\oplus_{\lambda \in \mathcal{S}_{\tiny \ydiagram{2}}} V^{\prime}_{\lambda}  $.  
In particular, 
 $V^{\prime}_{\lambda}$ coincides with the irreducible
  module  $V_{\lambda}$ 
 at least for $s=0$, $\lambda_{r}=0$, and
  $V^{\prime}_{\lambda} \simeq V^{+}_{\lambda} \oplus V^{-}_{\lambda}$ 
   at least for $s=0$, $\lambda_{r} \ne 0$, and then \eqref{de-t3D} coincides with the 
character formula of the Kirillov-Reshetikhin module $W_{a,m}=W^{(a)}_{m}$ for $U_{q}(\mathfrak{sl}(2r)^{(2)})$
 [eq.\ (6.9), \cite{HKOTT01}]
 (eq.\ \eqref{de-t3D} yields \eqref{de-t3D0} and \eqref{de-t3D00}
  upon substituting $s = 0$).  
The second equality \eqref{de-t3C} for $s=0$ corresponds to [eq.\ (6.8), \cite{HKOTT01}]
(eq.\ \eqref{de-t3C} yields \eqref{de-t3C0} and \eqref{de-t3D00}
  upon substituting $s = 0$), 
which suggests another decomposition of $W^{(a)}_{m}$. 

%%%%%%%%%%%%%%%%%%%%%%%%%%%%%%%%%%%%%%%%
\paragraph{$U_{q}(\mathfrak{osp}(2r|2s)^{(1)})$ case:}
Specializing the sets $X$ and $Y$ in \eqref{genh} 
to  finite sets: 
$X=\mathbf{x}\sqcup \mathbf{x}^{-1}$, 
$Y=\mathbf{y}\sqcup \{1,-1\} \sqcup  \mathbf{y}^{-1}$,  
we obtain
\begin{multline}
\prod_{j=1}^{r} (1-x_{j} t)^{-1} (1-x_{j}^{-1} t)^{-1} \prod_{j=1}^{s} (1-y_{j} t)(1-y_{j}^{-1} t)(1-t^{2})
=
\\
=\sum_{m=0}^{\infty} h_{m}(\mathbf{x} \sqcup \mathbf{x}^{-1}|\mathbf{y}\sqcup \{1,-1\} \sqcup \mathbf{y}^{-1}) t^{m} .
\label{gench-D}
\end{multline}
This corresponds to the folding of the $\mathfrak{gl}(2r|2s+2)$  case in  \eqref{gench-A}, under the identifications 
$x_{j}=z_{j}=z_{2r+1-j}^{-1}$ for $1 \le j \le r$, 
$y_{j}=z_{2r+j}=z_{2r+2s+3-j}^{-1}$ for $1 \le j \le s$ and  
$z_{2r+s+1}=-z_{2r+s+2}^{-1}=1$. 
Applying \eqref{dc-4} to $S_{\lambda}(X|Y)$, we obtain 
\begin{align}
S_{\lambda}(\mathbf{x}\sqcup \mathbf{x}^{-1}|\mathbf{y}\sqcup \{1,-1\} \sqcup  \mathbf{y}^{-1})&=\sum_{\mu \in {\mathcal P}}
\sum_{\kappa \in {\mathcal P}^{-} } \mathrm{LR}^{\lambda}_{\kappa , \mu} 
S_{[\mu ]}(\mathbf{x}\sqcup \mathbf{x}^{-1} |\mathbf{y}\sqcup   \mathbf{y}^{-1}).
\label{dc-4D}
\end{align}
Note that this is non-trivial only if $\lambda$ is in the 
$[2r,2s+2]$-hook. 
Specializing \eqref{dc-4D} to a
 rectangular Young diagram $(m^{a})$ in the $[2r,2s+2]$-hook ($a,m \in \mathbb{Z}_{\ge 1}$), and using \eqref{LRrq2}, we obtain the following decomposition formulas:
\begin{align}
\ytableausetup{boxsize=0.5em}
S_{(m^{a})}(\mathbf{x} \sqcup \mathbf{x}^{-1}|\mathbf{y}\sqcup \{1,-1\} \sqcup \mathbf{y}^{-1})
&=\sum_{\lambda \in \mathcal{S}_{\tiny \ydiagram{1,1}}} S_{[\lambda]} (\mathbf{x} \sqcup \mathbf{x}^{-1}|\mathbf{y}\sqcup \mathbf{y}^{-1}) 
&& \text{[on type $\Ds$]} .
\label{de-D}
\end{align}
To relate these formulas to the labels of representations, we  restrict them to the $[r,s]$-hook. 
In this case, our observation in \cite{T23} on the 
Bethe strap \cite{KS94-1,Su95} suggests that they do not necessarily yield irreducible supercharacters of $U_{q}(\mathfrak{osp}(2r|2s)^{(1)})$ in general, although they do at least in the case of the fundamental representation
 $\lambda =(1)$.
 In addition, \eqref{de-D} suggests a decomposition
\footnote{Compare this with eq.~(3.26) in \cite{T99-2} for the case $r=1$.}
  of a module $W_{a,m}$ of $U_{q}(\mathfrak{osp}(2r|2s)^{(1)})$ 
  (or $Y(\mathfrak{osp}(2r|2s))$)  
into modules $\{V^{\prime}_{\lambda}\}$ of $U_{q}(\mathfrak{osp}(2r|2s))$ (or $\mathfrak{osp}(2r|2s)$): $W_{a,m} \simeq
\oplus_{\lambda \in \mathcal{S}_{\tiny \ydiagram{1,1}}} V^{\prime}_{\lambda}  $.  
In particular, 
 $V^{\prime}_{\lambda}$ coincides with the irreducible module  $V_{\lambda}$ at least for $s=0$, $\lambda_{r}=0$, and
  $V^{\prime}_{\lambda} \simeq V^{+}_{\lambda}
   \oplus V^{-}_{\lambda}$ 
  at least for $s=0$, $\lambda_{r} \ne 0$, and then \eqref{de-D} 
  for $1 \le a \le r-2$ coincides with the 
character formula of the Kirillov-Reshetikhin module  
$W_{a,m}=W^{(a)}_{m}$ for $U_{q}(\mathfrak{so}(2r)^{(1)})$ or $Y(\mathfrak{so}(2r))$
 \cite{KR90} (see also [eq.\ (7.4), \cite{HKOTY98}]; 
 eq.\ \eqref{de-D} yields \eqref{de-D0} and \eqref{de-D00}
  upon substituting $s = 0$)
\footnote{
In Appendix~B of \cite{T23}, the highest weight module
$V(\Lambda)$ with highest weight $\Lambda$, viewed as a function of a partition (or Young diagram) $\lambda$, is also denoted by $V(\lambda)$.
This notation corresponds to $V_{\lambda}$ or $V^{\pm}_{\lambda}$ in the present paper.
We also point out that 
$V({\lambda}) \oplus V(\lambda)|_{\lambda_{r} \to -\lambda_{r}}$ in [page 63, \cite{T23}] is a misprint of 
$V({\lambda}) \oplus V(\lambda)|_{\Lambda_{r} \to -\Lambda_{r}}$, and  
$\chi_{\langle \mu \rangle}(\mathbf{x},\emptyset)$ in [page 61, \cite{T23}] is a misprint of 
$\chi_{\langle \mu \rangle}(\mathbf{x},\mathbf{x}^{-1}|\emptyset)$.
We do not impose the $\mathfrak{sl}(M|N)$-type constraint 
$z_{1}\cdots z_{M}\, z_{M+1}^{-1}\cdots z_{M+N}^{-1}=1$ 
\emph{\`{a} priori} in \eqref{gench-A}, although this relation holds
for the supercharacters of the twisted quantum affine superalgebras
after the folding procedures.
For this reason, in the present paper we use the notation
$\mathfrak{sl}(\dots)^{(2)}$ rather than $\mathfrak{gl}(\dots)^{(2)}$, even though the latter
was used in our previous paper~\cite{T23}.
In any case, this notational choice is inessential, since
representations of $\mathfrak{sl}(\dots)$ are naturally understood as those of
$\mathfrak{gl}(\dots)$.
}
 . The cases $a=r-1,r$ for $s=0$ have to be treated separately. 
%%%%

The algebra $U_{q}(\mathfrak{osp}(2r|2s)^{(1)})$ reduces to 
$U_{q}(\mathfrak{sp}(2s)^{(1)})$ when $r=0$.
To understand the relation between \eqref{de-C0}, \eqref{de-C00}
 and \eqref{de-D}, 
% one may view the Young diagrams in Figure 
%\ref{MN-hookD} after $90$-degree rotation
one may interchange the roles of rows and columns
in the Young diagrams in Figure \ref{MN-hookD}, in conjunction with   
\eqref{BCD-dual} and \eqref{A-dual}. 
Alternatively, 
we may specialize \eqref{dc-4du} as 
\begin{align}
S_{\lambda}(\mathbf{x} \sqcup \{1,-1\}\sqcup \mathbf{x}^{-1}|\mathbf{y} \sqcup \mathbf{y}^{-1})&=\sum_{\mu \in {\mathcal P}}
\sum_{\kappa \in {\mathcal P}^{+} } \mathrm{LR}^{\lambda}_{\kappa , \mu} 
S_{\langle \mu \rangle}(\mathbf{x} \sqcup \mathbf{x}^{-1} |\mathbf{y} \sqcup \mathbf{y}^{-1}),
\label{de-4du}
\end{align}
and, in particular, for a
 rectangular Young diagram $(m^{a})$ in the $[2r+2,2s]$-hook (with $a,m \in \mathbb{Z}_{\ge 1}$), and using \eqref{LRrq3}, 
\begin{align}
\ytableausetup{boxsize=0.5em}
S_{(m^{a})}(\mathbf{x} \sqcup \{1,-1\}\sqcup \mathbf{x}^{-1}|\mathbf{y} \sqcup \mathbf{y}^{-1})
&=\sum_{\lambda \in \mathcal{S}_{\tiny \ydiagram{2}}} S_{\langle \lambda \rangle} (\mathbf{x} \sqcup \mathbf{x}^{-1}|\mathbf{y}\sqcup \mathbf{y}^{-1}) 
&& \text{[on type $\Cs$]} .
\label{de-C}
\end{align}
Setting $s=0$ (so $\mathbf{y}\sqcup \mathbf{y}^{-1}=\emptyset$)
 reproduces \eqref{de-C0} and \eqref{de-C00}. 
 Eqs.\ \eqref{de-4du} and \eqref{de-C} are the  
 $U_{q}(\mathfrak{spo}(2r|2s)^{(1)})$ analogues of \eqref{dc-4D} and \eqref{de-D}, respectively. 
 %%%%%%%%%%%%%%%%%%%%%%%%%%%%%%%%%%%%%%%%
\paragraph{$U_{q}(\mathfrak{osp}(2r|2s)^{(2)})$, $r \ge 1$, $s\ge 0$ case:}
Specializing the sets $X$ and $Y$ in \eqref{genh} 
to  finite sets: 
$X=\tilde{\mathbf{x}}\sqcup \{1,1\} \sqcup \tilde{\mathbf{x}}^{-1}$, 
$Y=\mathbf{y}\sqcup \{1,-1\} \sqcup  \mathbf{y}^{-1}$,  
we obtain
\begin{multline}
\prod_{j=1}^{r-1} (1-x_{j} t)^{-1} (1-x_{j}^{-1} t)^{-1} \prod_{j=1}^{s} (1-y_{j} t)(1-y_{j}^{-1} t)(1- t)^{-1}(1+t)
=
\\
=\sum_{m=0}^{\infty} h_{m}(\tilde{\mathbf{x}}\sqcup \{1,1\}\sqcup \tilde{\mathbf{x}}^{-1}|\mathbf{y}\sqcup \{1,-1\} \sqcup \mathbf{y}^{-1}) t^{m} ,
\label{gench-D2}
\end{multline}
where  
$\tilde{\mathbf{x}}=\{x_{j} \}_{j=1}^{r-1}$, $\tilde{\mathbf{x}}^{-1}=\{x_{j}^{-1} \}_{j=1}^{r-1}$. 
This corresponds to the folding of the $\mathfrak{gl}(2r|2s+2)$  case in  \eqref{gench-A}, under the identifications 
$x_{j}=z_{j}=z_{2r+1-j}^{-1}$ for $1 \le j \le r-1$, 
$z_{r}=z_{r+1}^{-1}=1$, 
$y_{j}=z_{2r+j}=z_{2r+2s+3-j}^{-1}$ for $1 \le j \le s$ and  
$z_{2r+s+1}=-z_{2r+s+2}^{-1}=1$. 
Applying \eqref{SP-A} and \eqref{dc-2b} for $\xi =-1$ to $S_{\lambda}(X|Y)$, we obtain 
\begin{align}
S_{\lambda}(\tilde{\mathbf{x}}\sqcup \{1,1\} \sqcup \tilde{\mathbf{x}}^{-1} |\mathbf{y}\sqcup \{1,-1\} \sqcup  \mathbf{y}^{-1})
&=
S_{\lambda}(\tilde{\mathbf{x}}\sqcup \{1\} \sqcup \tilde{\mathbf{x}}^{-1} |\mathbf{y}\sqcup \{-1\} \sqcup  \mathbf{y}^{-1})
\nonumber \\
&=\sum_{\mu ,\nu \in {\mathcal P}}
 \mathrm{LR}^{\lambda}_{\nu , \mu} 
S_{[\mu]}(\tilde{\mathbf{x}}\sqcup \{1\} \sqcup \tilde{\mathbf{x}}^{-1} |\mathbf{y}\sqcup  \mathbf{y}^{-1} ).
\label{dc-5cD2}
\end{align}
Note that this is non-trivial only if $\lambda$ is in the 
$[2r,2s+2]$-hook. 
Specializing \eqref{dc-5cD2} to a
 rectangular Young diagram $(m^{a})$ in the $[2r,2s+2]$-hook ($a,m \in \mathbb{Z}_{\ge 1}$), we obtain the following decomposition formulas:
\begin{align}
\ytableausetup{boxsize=0.5em}
S_{(m^{a})}(\tilde{\mathbf{x}}\sqcup \{1\}\sqcup \tilde{\mathbf{x}}^{-1}|\mathbf{y}\sqcup \{-1\}\sqcup \mathbf{y}^{-1})
&=\sum_{\lambda \in \mathcal{S}_{\tiny \ydiagram{1}}} S_{[\lambda]} (\tilde{\mathbf{x}}\sqcup \{1\}\sqcup \tilde{\mathbf{x}}^{-1}|\mathbf{y} \sqcup \mathbf{y}^{-1}) 
&& \text{[on type $\Bs$]} .
\label{de-D2}
\end{align}
To relate these formulas to the labels of representations, we  restrict them to the $[r-1,s]$-hook. 
%In this case, our observation on the Bethe strap suggests that  it does not always give an
% irreducible character of $U_{q}(\mathfrak{osp}(2r|2s)^{(2)})$ in the general situation. 
% In addition, 
\eqref{de-D2} suggests a decomposition of a module $W_{a,m}$ of $U_{q}(\mathfrak{osp}(2r|2s)^{(2)})$ 
into modules $\{V^{\prime}_{\lambda }\}$ of $U_{q}(\mathfrak{osp}(2r-1|2s))$: $W_{a,m} \simeq
\oplus_{\lambda \in \mathcal{S}_{\tiny \ydiagram{1}}} V^{\prime}_{\lambda}  $.  
In particular, 
 $V^{\prime}_{\lambda}$ coincides with the irreducible module  $V_{\lambda}$ 
 at least for $s=0$, $\lambda_{r-1}=0$, and then \eqref{de-D2} 
   for $1 \le a \le r-2$ coincides with the 
character formula of the Kirillov-Reshetikhin modules $W_{a,m}=W^{(a)}_{m}$ for $U_{q}(\mathfrak{so}(2r)^{(2)})$ 
 [eq.\ (6.10), \cite{HKOTT01}]
 (eq.\ \eqref{de-D2} yields \eqref{de-D20} and \eqref{de-D200}
  upon substituting $r \to r+1$ and $s = 0$). 
 The case $a=r-1$ for $s=0$ has to be treated separately. 
 
In closing this section, we remark that Weyl-type determinant formulas (that is, formulas expressed as ratios of determinants, such as the standard determinantal expression \eqref{SchDet} for Schur functions) for quantum affine (super)algebras can also be obtained by applying the same reduction procedures described in this section to the determinant formulas for (super)characters of 
$\mathfrak{gl}(M|N)$ given in \cite{MV03}, instead of to the 
Jacobi-Trudi-type formulas. We further expect that these reduction  procedures extend to a broader family of (super)symmetric polynomials or functions, not limited to the supersymmetric Schur functions treated in this work. 
 
%%%%%%%%%%%%%%%%%%%%%%%%%%%%%%%%%%%%%%%%

\section{Conclusion}

In this paper we have established that the characters of Kirillov-Reshetikhin (KR) modules for quantum affine algebras can be obtained as folded supercharacters of $\mathfrak{gl}(M|N)$. 
This result confirms the conjecture formulated in our earlier work \cite{T23} and demonstrates that the folding procedure, together with Cauchy-type relations for $\mathfrak{gl}(M|N)$ and $\mathfrak{osp}(M|N)$ supercharacters, provides an efficient and algebraically transparent method for deriving explicit formulas for KR characters. 
This framework offers a unified perspective connecting various classes of quantum affine algebras-both twisted and untwisted-with the representation theory of quantum affine superalgebras. 
In particular, the correspondence established here elucidates structural parallels between bosonic and supersymmetric settings and clarifies how supercharacters encode information extending beyond the purely type $A$ case. 

Our results further encompass the case of quantum affine orthosymplectic superalgebras, or twisted quantum affine superalgebras, where the supercharacters admit decomposition formulas in terms of those of finite-type super-subalgebras.  
A crucial distinction from the non-superalgebraic case, however, is that such decompositions do not necessarily yield irreducible supercharacters in general, particularly for the 
$U_{q}(\mathfrak{osp}(2r|2s)^{(1)})$ case.  
It is plausible that certain modifications, such as subtracting the contributions from invariant subspaces, are required to isolate the irreducible components.  
Establishing a systematic procedure for such refinements remains an open problem.  
Although a general criterion for irreducibility is not yet known, insight into this issue may be obtained through the so-called \emph{Bethe strap} structure \cite{KS94-1,Su95}, which arises in the analysis of the Bethe ansatz.  
As discussed in our previous work \cite{T23}, the Bethe strap reveals a combinatorial organization of modules that may serve as a guiding principle for understanding the irreducibility of supercharacters. 

In addition to the above issue, several further directions remain open for future development. These include the explicit construction of representations of non-type A quantum affine (super)algebras derived from the evaluation representations of
 $U_{q}(\mathfrak{gl}(M|N)^{(1)})$, as well as their applications to quantum integrable systems-in particular, the explicit formulation of the corresponding R-matrices and the associated T- and Q-operators. 
Pursuing these directions is expected to further elucidate, within a unified framework, the interplay between the representation theory of quantum affine (super)algebras and the analytic structures underlying quantum integrable models.

%%%%%%%%%%%%%%%
\section*{Acknowledgments} 
The work was partially supported by Grant No. 0657-2020-0015 from the Ministry of Science and Higher Education of Russia, awarded to the
Pacific Quantum Center at Far Eastern Federal University, Vladivostok, Russia. 
I would like to thank the referee for the helpful comments on the manuscript.
%%%%%%%%%%%%%%%%%%%%
\section*{
Declaration of generative AI and AI-assisted technologies in the manuscript preparation process}
During the preparation of this work the author used ChatGPT and Gemini (free versions) in order to improve the language and flow of the Abstract, Section 1 and Section 5. After using these tools, the author reviewed and edited the content as needed and takes full responsibility for the content of the published article.
%%%%%%%%%%%

\end{document}